\documentclass[journal]{IEEEtran}
\usepackage{placeins}
\usepackage{graphicx}
\usepackage{caption}
\usepackage{subcaption}
\usepackage{bbold}
\usepackage{gensymb}
\usepackage{xcolor}
\usepackage [acronym] {glossaries}
\newacronym{design}{Panoptic}{Rx Beamforming for Joint Communication and Sensing}

\usepackage[normalem]{ulem}
\usepackage{xcolor}

\begin{document}
%
\title{Panoptic: True Joint mmWave Communication and
Sensing with Compressive Sidelobe Forming}
%
%
\author{Heyu~Guo,
        Ruiyi~Shen, Florian Kosterhon, and
        Yasaman~Ghasempour
\thanks{H. Guo, R. Shen, and Y. Ghasempour are with the Department
of Electrical and Computer Engineering, Princeton University, Princeton,
NJ, 08644 USA e-mail: \{hg9046, ruiyishen, ghasempour\}@princeton.edu. F. Kosterhon is with TU Delft e-mail: F.Kosterhon@tudelft.nl. This work was conducted while F. Kosterhon was affiliated with Princeton University.}
\thanks{© 2025 IEEE. Personal use of this material is permitted. Permission
from IEEE must be obtained for all other uses, in any current or future
media, including reprinting/republishing this material for advertising or
promotional purposes, creating new collective works, for resale or
redistribution to servers or lists, or reuse of any copyrighted
component of this work in other works.}}

\markboth{Journal on Selected Areas in Communications, 2025}%
{Shell \MakeLowercase{\textit{et al.}}: Panoptic: True Joint mmWave Communication and
Sensing with Compressive Sidelobe Forming}

\maketitle

\begin{abstract}
The integration of communication and sensing functions within mmWave systems has gained attention due to the potential for enhanced passive sensing and improved communication reliability. State-of-the-art techniques separate these two functions in frequency, use of hardware, or time, i.e., sending known preambles for channel sensing or unknown symbols for communications. In this paper, we introduce \acrshort{design}, a novel system architecture for integrated communication and sensing sharing the same hardware, frequency, and time resources. \acrshort{design} jointly detects unknown symbols and channel components from data-modulated signals. The core idea is a new beam manipulation technique, which we call compressive sidelobe forming, that maintains a directional mainlobe toward the intended communication nodes while acquiring unique spatial information through pseudorandom sidelobe perturbations. We implemented \acrshort{design} on 60 GHz mmWave radios and conducted extensive over-the-air experiments. Our results show that \acrshort{design} achieves reflector angular localization error of less than $2^\circ$ while at the same time supporting mmWave data communication with a negligible BER penalty when compared with conventional communication-only mmWave systems. 
\end{abstract}


\IEEEpeerreviewmaketitle

\section{Introduction} \label{sec:intro}

Millimeter-wave (mmWave) wireless networks are emerging thanks to the wide available bandwidth and multi-Gbps data rates that enable new applications like extended reality~\cite{chakareski2023millimeter}, and high-quality video streaming~\cite{zhang2021innovating}, among other applications. On the other hand, GHz-scale bandwidth, mm-scale wavelength, and the ability to manipulate EM radiation using large antenna arrays have made this spectral regime attractive for localization, sensing, and even imaging tasks~\cite{liu2022mtranssee, liu2020real, zhang2022synthesized, pefkianakis2018accurate, adhikari2022mishape, regmi2021squigglemilli}. The convergence of sensing and communication functions in mmWave wireless networks is a promising trend that has gained increasing momentum in recent research and industrial developments~\cite{paidimarri2024eye,li2024gemini}. Indeed, such an interaction enables two key advantages over traditional wireless systems: First, it can offer additional services from existing communication resources and infrastructure, and second, the sensing input from the surrounding medium can significantly enhance the reliability of mobile mmWave communications. \textcolor{blue}{}

The state-of-the-art integrated communication and sensing techniques separate the two functionalities either in time or frequency, or exploit two distinct hardware (e.g., mmWave radar and communication module)~\cite{mollahosseini2025integrated, pegoraro2024jump, alrabeiah2020deep, graff2023deep}. Therefore, despite promising prospects, these efforts do not achieve true joint communication and sensing functionalities\footnote{We refer to sensing as finding and tracking surrounding reflecting objects in real-time, which for example can form  back-up paths in case of blockage.} that share time, frequency, and hardware resources. 

In this paper, we present \acrshort{design}\footnote{In Greek mythology, Argus Panoptes is a giant with 100 eyes, whose name means ``all-seeing".}, a novel framework for integrated communication and sensing at the same time, and frequency band. \acrshort{design} does not require additional hardware deployment (e.g., radar) and can be implemented using pervasive and widely available single-RF chain mmWave antenna arrays. The fundamental challenge behind realizing this goal is two-fold: \textit{(i)} the communication and sensing functions have inherently different requirements, i.e., sensing needs time-varying directional beam scanning while communication demands a stable and accurately aligned directional beam to achieve high SNR; \textit{(ii)} communication includes detection of \textit{unknown} data symbols while sensing involves analyzing the properties of \textit{known} signals (e.g., cross-correlation of received and known preambles) to extract the key components of the wireless medium.

\begin{figure}[t]
  \centering
  \includegraphics[width=\linewidth]{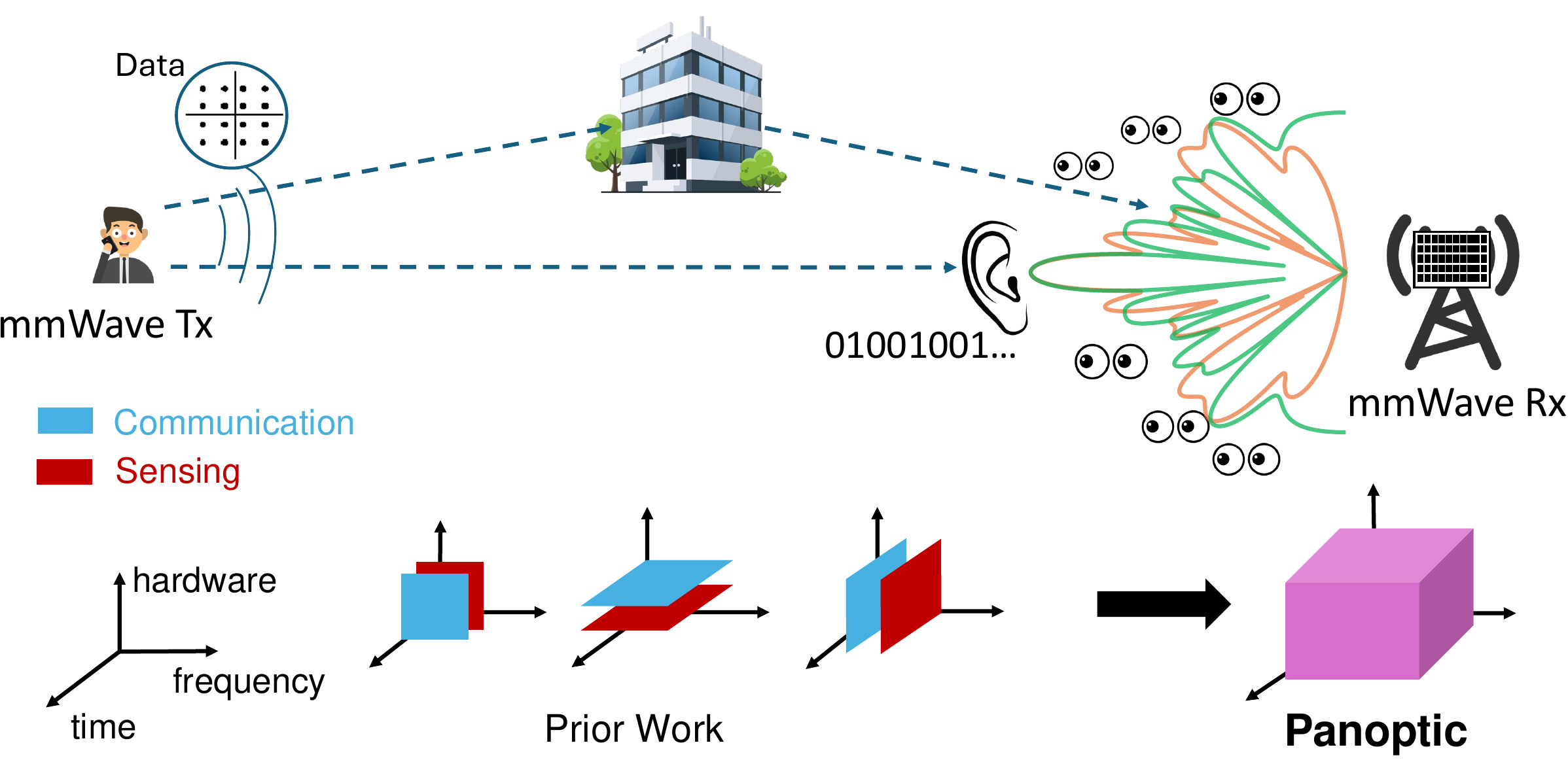}
   \vspace{-3mm}
  \caption{Overview of \acrshort{design}, a framework for joint data reception and reflector sensing at the same time, frequency, and using the same hardware resources.}
  \label{fig:overview}
  \vspace{-4mm}
\end{figure}

\acrshort{design} tackles these challenges by introducing a novel antenna modulation and compressive signal processing framework that capitalizes on channel sparsity. Specifically, our key idea is to maintain a directional mainlobe toward the communication node while simultaneously generating pseudorandom sidelobe perturbations to allow for reflection sensing in non-mainlobe directions. The high-level overview of \acrshort{design} is depicted in Fig.~\ref{fig:overview}  illustrating a bistatic integrated communication and sensing architecture. A mmWave data-carrying link is formed between the two communication parties while, at the same time, sidelobe perturbations on one end assist with environmental tracking.  Specifically, a reflected path yields random, albeit unique power and phase fluctuations at the receiver for a given suite of randomly shaped sidelobes. Such variations may act as a signature for path sensing when properly extracted from the overall superposed signals containing unknown complex-valued symbols. 

To realize such desired beam manipulation, \acrshort{design} leverages antenna subset modulation. Particularly, we randomly turn off (or change the gain of) a subset of antennas in each round. Since the relative phase delays among the antennas are unchanged, the remaining ON antennas still
form constructive interference along the same mainlobe directions while random destructive interference perturbs the radiation gain in all other angles. Indeed, our compressive sidelobe architecture enables separating the sensing and communication functionalities in space, while its implementation requires simple single-chain arrays with antenna ON/OFF switching that are commercially available and even already deployed in many real-world settings.

While our compressive sidelobe forming framework addresses challenge \textit{(i)} explained above, the problem of joint data recovery and path sensing has remained unsolved. Specifically, the received signal is a combination of an unknown data-modulated signal that travels through a set of unknown paths. Further, the non-line-of-sight (NLOS) signatures are inherently weaker due to additional path loss and significantly smaller directivity gain along the sidelobe directions where such paths are perceived. Hence, the NLOS fingerprints may
be masked under a much stronger mainlobe signal. To tackle this, we mathematically model the received signal and present a new signal processing framework that performs consecutive symbol detection and then NLOS angle estimation by integrating information across several compressive sidelobe patterns. Therefore, \acrshort{design} enables real-time and continuous reflector identification and tracking at zero additional overhead, i.e., through joint sensing and data reception in mobile environments. Indeed, such environmental inputs are the key to immediate link recovery under blockage, and realizing reliable and resilient mmWave networking.

We have implemented \acrshort{design} on off-the-shelf mmWave arrays and performed extensive over-the-air experiments at 60 GHz in various configurations in two different environments: a conference room and a large open lab space. As a baseline, we implemented conventional mmWave directional beamforming, in which all antennas are constructively forming an aligned beam toward the communication node; thus, maximizing the link SNR with no sensing capability. Our experimental results demonstrate that \acrshort{design} provides an accurate estimation of NLOS angle with a mean error of less than $1.3\degree$ and the standard deviation of less than $1.2\degree$ while incurring a negligible penalty on BER performance when compared with the baseline scheme under different modulation schemes. Our results also suggest that \acrshort{design} is scalable to detecting multiple reflectors in the medium as the average NLOS angle estimation error only slightly rises to 1.70$\degree$ with two reflectors. 

In addition, we have also experimentally investigated the tradeoff between communication and sensing functionalities. Specifically, by randomly turning off a larger subset of antennas, the sidelobe randomness space grows and better sensing accuracy is observed, albeit at the cost of a higher SNR reduction for  the mainlobe communication link. Fortunately, this penalty alleviates when large antenna arrays are employed. Indeed, the length of the NLOS perturbation signature correlates with $N\choose k$, where $k$ is the number of OFF antennas out of the total $N$ antennas while the mainlobe directivity scales by a factor of $1-k/N$.  Interestingly, our results reveal that randomly turning off \textcolor{black}{only 2 out of 64 antenna elements} creates sufficiently large random space for accurate NLOS angle estimation (error of less than 1.5 degrees) while penalizing the communication SNR by only a negligible amount of \textcolor{black}{0.28 dB} (which does not often result in  downgrading the Modulation and Coding Scheme (MCS) of the communication link).
Further, thanks to its compressive sensing nature, the number of required beam configurations only grows by $O({\rm log}(N))$ with the number of antennas; therefore, \acrshort{design} is scalable to future massive mmWave arrays. Finally, \acrshort{design} is compliant with IEEE 802.11ay and 5G NR as it does not impose any changes to the communication protocol and frame structure.

\section{Primer and State-of-the-Art} \label{sec:background}

\subsection{Principles of Joint Communication and Sensing}

Recently, developing joint communication and sensing (JCS) techniques have emerged to integrate the two functions into one by sharing hardware and signal processing modules, and achieving immediate benefits of reduced cost, size, weight, and better spectral efficiency~\cite{zhang2021enabling}. In wireless communication networks, the received signal is a function of both the unknown data symbols and the propagation medium, which we intend to sense. In particular, we can write: 
\begin{equation}
y = \mathbf{W_r}^* \mathbf{H} \mathbf{W_t} s + n,
\label{equ: received-signal}
\end{equation}
where $\mathbf{W_r}$ is the receiver weight vector, * is conjugate transpose, $\mathbf{H}$ is the channel response, $\mathbf{W_t}$ is the transmitter weight vector, $s$ is the data symbol and $n$ is the noise. The goal of wireless sensing is to extract $\mathbf{H}$, which carries key environmental components. However, from Eq.~\eqref{equ: received-signal}, it is evident that simultaneous channel estimation ($\mathbf{H}$) and data decoding ($s$) is not trivial.

Therefore, most existing efforts in JCS leverage \textit{known preambles and pilots} that are already part of the IEEE 802.11 or 5G NR frame structures for sensing. Existing mmWave sensing systems have repurposed the beam training period in IEEE 802.11ad/ay, during which the AP adopts sequential directional beams toward different angles.  By analyzing the received sector sweep frames, one can then track the channel multipath or the angular location of the transmitting or receiving node passively ~\cite{lacruz2020mm, jog2019many, ghasempour2018multi,ghasempour2019multi, hassanieh2018fast,garcia2020polar, shen2024characterizing}.  However, the beam training period is a low-duty cycle event (e.g., every 100 ms) originally designed to find the best-aligned beams between the AP and stations.  Similarly, in 5G NR protocol, the SSB bursts can be sent with a periodicity of 5, 10, 20, 40, 80, or 160 ms (where typically SSB bursts take 5 ms in duration) and leveraged for sensing~\cite{jain2021two,paidimarri2024eye}. Clearly, a more frequent SSB burst can offer advantages for sensing at the cost of increased overhead (i.e., occupying channel resources that be otherwise used for data communication). We emphasize that, unfortunately, these techniques do not support truly joint communication and sensing: when the AP/BS is performing beam training, it can sense the environment but without data communication; when the AP and stations are communicating, sensing cannot be achieved.

In addition, the preamble and header of a directional multi-gigabit (DMG) packet in IEEE 802.11ad can be used for sensing{~\cite{pegoraro2022sparcs, 80211ad}.  However, according to the standard, the preamble, header, and data fields of a DMG packet are transmitted with the same antenna weight vector (AWV), which limits their capability to sense channel dynamics (e.g., reflectors) outside of the mainlobe. 


In contrast, \acrshort{design} supports integrated directional data reception and quasi-omni sensing without adding any additional overhead using unknown data symbols.


\begin{figure*}[t]
    \centering
    \includegraphics[width=0.8\textwidth]{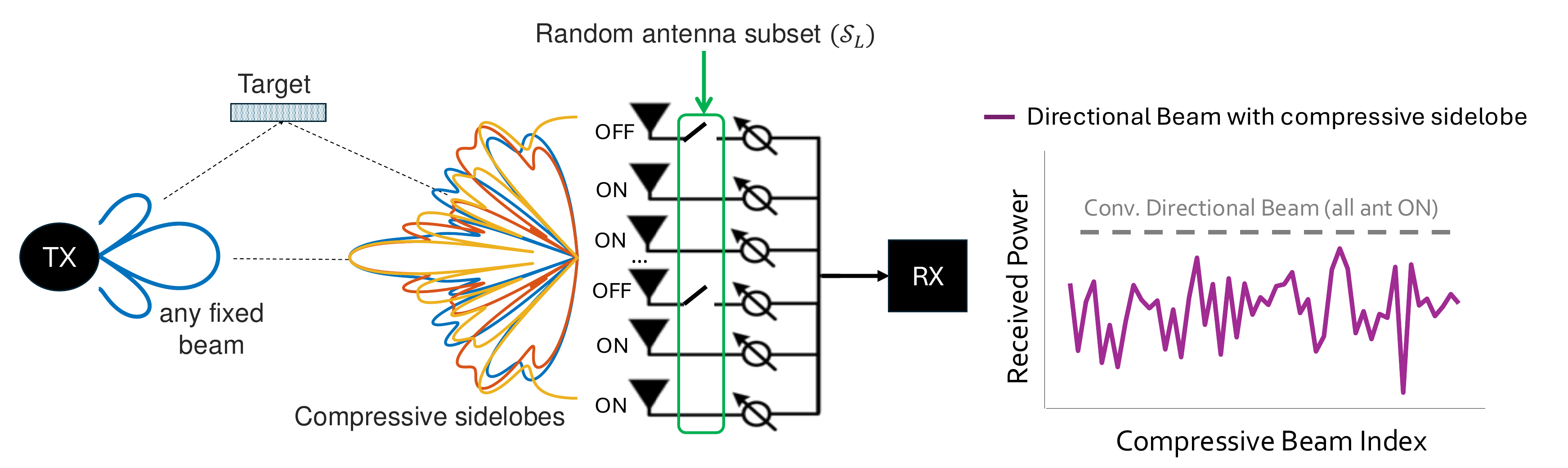}
    \vspace{-2mm}
    \caption{The architecture for compressive sidelobe beamforming that maintains a directional and aligned mainlobe for communication and creates pseudorandom sidelobes for NLOS sensing.}
    \label{fig:design}
     \vspace{-3mm}
\end{figure*}

\subsection{JCS with Auxiliary Spectral Bands or Hardware}
To achieve communication and target sensing at the same time, some prior efforts mount additional hardware on the mmWave base station or the mobile users to act as a passive radio sensor or an active transceiver, albeit at a different frequency band to avoid interference on the ongoing communications. This includes the use of co-located sub-6 GHz radios~\cite{alrabeiah2020deep}, mmWave radar \cite{graff2023deep}, reflective surfaces \cite{jiao2021enabling}, cameras \cite{yang2023environment}, and the auxiliary mmWave radios \cite{guan20213, yu2023mmalert}.


However, these works require additional hardware or spectrum resources hindering spectrum efficiency. Mounting additional hardware increases power consumption and may even yield new privacy concerns (e.g., in the case of the camera). Further,  the cost of hardware in addition to the prohibitively high labor cost associated with implementation and maintenance make these solutions impractical and non-scalable beyond laboratory settings. In contrast, \acrshort{design} uses existing single-chain analog arrays prevalent in all mmWave nodes (base stations, access points, and mobile users) without the need for additional hardware, spectrum resources, or even without any hardware modification, making it easily scalable in real-world settings. 

\subsection{Multi-Beam Full Duplex JCS Systems}
Different from the above strategies, another line of work exploits spatial division between sensing and communication via beam manipulation~\cite{ barneto2020multibeam, barneto2022beamformer}. Although beamforming for communication and sensing separately has been studied extensively~\cite{xue2024survey}, its application in JCS comes with a different set of challenges. Specifically, communication and sensing have different beam requirements: sensing needs time-varying directional beam scanning while communication demands stable and accurately aligned directional beam to achieve high SNR. More recently, multi-beam designs using full-duplex radios have been proposed, where there is a fixed directional beam for downlink communication and the other sweeping beam for collecting the reflected signal back for sensing~\cite{yang2023side, tang2021self,kumari2018sparsity}. In such case, complex beamforming optimization is needed to mitigate the self-interference between communication and sensing at the full-duplex node~\cite{roberts2021hybrid, roberts2024real, roberts2019beamforming}. In such full-duplex monostatic architectures, the transmitted signals are known at the co-located receiver and can be used as a reference  to easily extract the desired reflection information from nearby objects. Overall, full duplex radios are not common in practical infrastructure settings, especially for mobile devices, because of the increased hardware complexity and power consumption.


\textcolor{black}{In contrast to the prior art, \acrshort{design} has fundamentally different hardware requirements and system goals: \textit{(i)}
\acrshort{design} employs half-duplex single-chain analog phased arrays that are commonly available in all mmWave nodes. Hence, there is no hardware barrier for deploying \acrshort{design} on existing infrastructure. \textit{(ii)} \acrshort{design} has a bistatic JCS architecture. It eliminates self-interference since transmission and reception occur at separate nodes. This makes it well-suited for deployment on commercially available, single-RF-chain mmWave arrays. Additionally, they align naturally with existing mobile network roles, supporting practical integration into real-world systems. Further, the bistatic JCS architecture in \acrshort{design} means that nodes might not be tightly synchronized posing additional challenges for sensing. \textit{(iii)} Finally, \acrshort{design} aims to extract the unknown data symbols jointly with environmental components. This is not a problem in full-duplex settings, as the data is already known at the co-located TX/RX node. \acrshort{design} is the first to jointly decode unknown data symbols and sense channel components at the same time.}

\section{Panoptic Design}  \label{sec:design}

\subsection{System Model} \label{subsec:degisn_overview}
We consider the case where two mmWave nodes have established a directional communication link. We assume each node is equipped with a conventional analog beamforming array connected to a single RF chain, i.e., an $N_{tx}$-element TX antenna array and $N_{rx}$-element RX antenna array. For simplicity, we assume that the antenna arrays are uniform linear arrays but extension to arbitrary antenna arrays is straightforward. Our goal is to jointly decode unknown data symbols while extracting the sparse channel components (i.e., reflected paths) at the same time, and frequency, and using the same one-chain receiver antenna array.

We present \acrshort{design}, a novel beam manipulation technique for true joint communication and sensing. The key idea is maintaining the receiver mainlobe toward the communication node while simultaneously generating pseudorandom perturbations on sidelobes. Thanks to the sparsity of mmWave channels and compressive sensing principles, \acrshort{design} is able to extract the target components in the environment (i.e., the angular location of reflectors) while correctly detecting the unknown sent symbols. 

The \acrshort{design} system model is shown in Fig.~\ref{fig:design} where the transmitter adopts any fixed beam while the receiver has time-varying compressive sidelobe patterns.  The pseudorandom perturbation of the sidelobes can be realized by randomly turning off (or changing the gain) a subset of antennas for each reception. Since the relative phase delays among the antennas are unchanged, the remaining antennas still form constructive interference along the same mainlobe directions while random destructive interference perturbs the radiation gain in all other angles. We will show that this simple scheme creates a wide angular field of view (FoV) for sensing. We underscore that such a radiation manipulation technique does not need hardware modification or multiple RF chains as almost all commodity phased arrays allow for at least 1-bit amplitude control (ON/OFF) per antenna. 

Another key challenge is simultaneously extracting symbols and target information from received data-modulated signals. This is non-trivial because the received signal contains an unknown combination of the transmitted symbols and the propagating environment. Further, the sensing sidelobes are much weaker than the communication mainlobe, making accurate and reliable sensing much more difficult. In other words, the sidelobe variations yield minute perturbations in the overall captured phase and amplitude at the RX, which acts as a unique fingerprint for the target's angular location. However, such information-rich perturbations may be masked under the data-induced phase/amplitude variations that are received under a much stronger mainlobe direction. This is demonstrated in Fig.~\ref{fig:algo_sensing}. To tackle this issue, we perform integration of multiple signals for sensing, where the integration window length can be adjusted subject to sidelobe power. Our key insight is that the antenna ON/OFF switching is controllable and can happen at a much lower pace than the symbol rate; hence, several samples are effectively collected under the same sidelobe pattern. Indeed, the sensing time scale follows the mobility time scale (at millisecond scale) and is slower than the symbol transmission rate (at MHz or even GHz). \textcolor{black}{Finally, \acrshort{design} provides a \textit{non-coherent} target sensing framework that only utilizes the amplitude of received signals, and therefore is robust to carrier frequency offset and synchronization non-idealities between the two communicating parties. }

Next, we will explain our theory of compressive sidelobe beams and present the joint data decoding and sensing algorithm in \acrshort{design}.

\begin{figure*}[h!]
    \centering
    \includegraphics[width=0.9\textwidth]{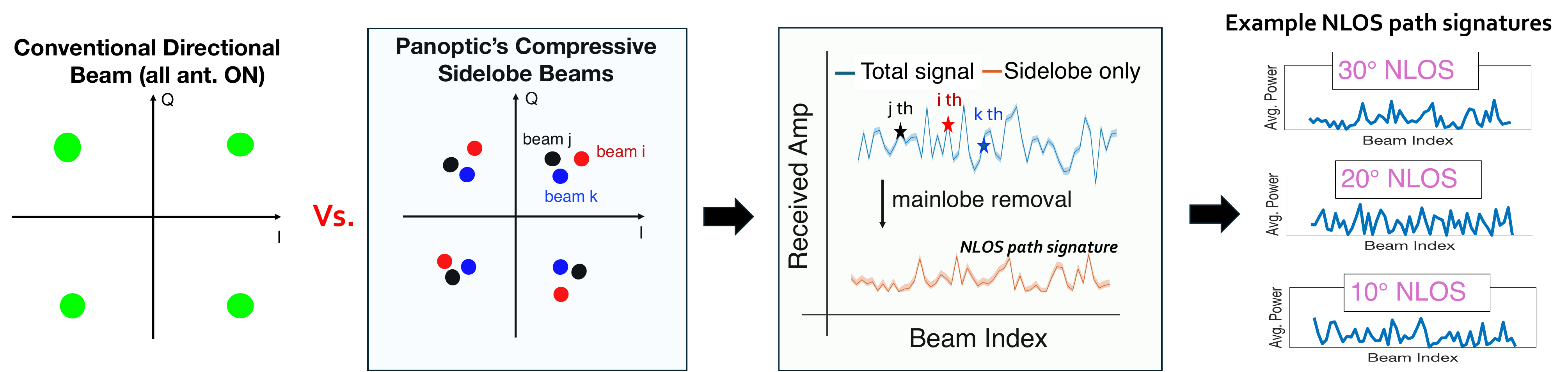}
    \vspace{-1mm}
    \caption{Compressive sidelobe beam creates small but pseudorandom perturbations on the received constellation, creating a unique NLOS-dependent fingerprint. Yet, such signatures might appear hidden under a stronger mainlobe.}
    \vspace{-3mm}
    \label{fig:algo_sensing}
    
\end{figure*}

\subsection{Compressive Sidelobe Beam Design} \label{subsec:css}
The objective of adaptive compressive sidelobe beam design is to ensure a directional communication link maximizing the beamforming gain between the transmitter and the target receiver (described by the angular location AoA (Angle of Arrival) / AoD (Angle of Departure) of $(\phi_0,\theta_0)$) while at the same time causing sufficient perturbations at other angles ($\phi \neq \phi_0$ and/or $\theta \neq \theta_0$) for NLOS sensing.
Such radiation perturbations can be created by manipulating the weight vector at the transmit/receive mmWave arrays.

We consider a sensing interval of $T$ seconds during which the channel variation is small enough to allow for a quasi-stationary assumption of channel paths and end users. Note that $T$ can be adapted as a control knob in \acrshort{design}'s framework. During this sensing interval, the mmWave end nodes adopt $M$ different beam configurations \textcolor{black}{, where the RX collects $N$ symbols under each beam configuration}. We can write the received signal for the $n^{th}$ symbol under the $m^{th}$ beam configuration as follows: 
\begin{equation}
y[n,m]=w_r^* [m] H w_t [m] s[n,m]+ z[n,m],
\label{eq:received_nm}
\end{equation}
where $w_r [m]$ and $w_t [m]$ are the $m^{th}$ receiver and transmitter weight vectors, respectively, \textcolor{black}{$s[n,m]$ is the symbol for each time slot,} and $z[n,m]$ is the zero-mean noise. The wireless channel $H$ consists of $K$ multipath components described by their AoA/AoD pair $(\phi_k,\theta_k)$ and complex channel gain $\alpha_k$.  Hence, we can write $H$ as follows:
\begin{equation}
H= \sqrt{\alpha_0} \textbf{b}(\phi_0) \textbf{a}^* (\theta_0)+ \sum_{k=1}^K \sqrt{\alpha_k} \textbf{b}(\phi_k) \textbf{a}^* (\theta_k),
\end{equation}
where $\textbf{b}(.)$ is the steering vector for the TX array and $\textbf{a}(.)$ is the steering vector for the RX array. We note that $K$ is small in practice due to the sparsity in mmWave channels.

In principle, the pseudorandom sidelobe perturbations can be realized in both TX arrays and RX arrays. However, in practice, changing the radiation profile on both sides at the same time may cause challenges in sensing and inference. This is because the weight vector variation on both ends (i.e., $w_r[m]$ and  $w_t [m]$ in Eq.~\eqref{eq:received_nm}) should be somewhat synchronized. In other words, the receiver should know the weight vector (i.e., sidelobe patterns) adopted at the TX for each signal reception. Hence, without loss of generality, we consider a fixed transmit beam $w_t[m] =w_t$ and a varying receive radiation pattern. 

We highlight that the \acrshort{design}'s framework can also be directly applied to a fixed received beam and varying transmit sidelobe beams. In Sec.~\ref{subsubsec:tradeoff}, we will illustrate the communication-sensing tradeoffs and discuss that it is desirable to implement \acrshort{design}'s compressive sidelobes on the mmWave end node with more antennas (whether TX or RX). \textcolor{black}{Furthermore, the fixed beam on one end is pointed toward the intended communication party. There is an inherent trade-off in selecting its beamwidth: a narrower beam improves communication SNR but restricts the spatial extent of illumination, limiting the sensing capability. In contrast, a wider beam increases the likelihood of illuminating nearby reflectors, enhancing the quality of sensing information, albeit at the cost of reduced SNR. This trade-off is well understood in the community and is independent of the proposed \acrshort{design} framework.}


To create pseudorandom perturbations on the sidelobes, \acrshort{design} uses a randomly chosen subset of antennas to form a directional received beam toward the TX while the remaining antennas are kept idle. Since the antenna amplitudes are changing (ON vs. OFF) and not their phases, this creates sidelobes without changing the mainlobe direction, as also depicted in Fig.~\ref{fig:design}. Mathematically, let $\mathcal{S}_L[m]$ be a random subset of $L$ antennas (out of $N_{rx}$) used for the $m^\text{th}$ pattern configuration. We can write the varying received weight vector with the mainlobe pointing toward the TX (i.e., $\phi_0$ AoA) as:

\vspace{-4mm}
\begin{equation}
w_r[m] (l)=
  \begin{cases}
      e^{j(l-1)2\pi \frac{d}{\lambda} cos (\phi_0)}, & \text{$l$ $\in$ $\mathcal{S}_L[m]$}\\
      0, & \text{otherwise}
    \end{cases}
    \label{eq:w_r}
\end{equation}
\vspace{-2mm}

\noindent where $w_r[m] (l)$ is the $l^{\text{th}}$ entry of the received weight vector (i.e., $l=1:N_{rx}$) and $d$ is the antenna spacing. Hence, by combining the steering vector and the weight vector from Eq.~(\ref{eq:w_r}), we can find the array factor (or radiation profile) of the receiver beam as follows:

\vspace{-2mm}
\begin{equation}
R(m,\phi)=\sum_{\text{$l$ $\in$ $\mathcal{S}_L[m]$}} e^{j(l-1)2\pi \frac{d}{\lambda} (cos \phi_0 -cos\phi)}.
\label{eq:radiationP}
\end{equation}
\vspace{-2mm}

\noindent We highlight three important observations: First, $R(m,\phi)$ converges to a constant $R(m,\phi)=L$ at the target communication transmitter location ($\phi=\phi_0$). Indeed, all antennas being ON (i.e., $L=N_{rx}$) provides the maximum directivity gain toward the TX. Second, the radiation gain toward other angles ($\phi \neq \phi_0$) is a \textit{random variable} that is dependent on $\phi$, the antenna subset size $L$ and the exact subset of antennas that are ON. Finally, random sidelobe perturbations are realized solely with binary amplitude control, making it applicable to practical phased arrays that are not equipped with complex phase/amplitude control circuitry. 

\subsection{Joint Data Detection and NLOS Sensing Framework} \label{subsec:algo}
In this section, we explain how \acrshort{design} exploits compressive sidelobes in joint data detection and NLOS extraction. Recall from Sec.~\ref{subsec:css} that while symbols should be detected at the symbol rate, the sensing time-scale is slower as $M$ compressive sidelobe beams should be observed to reliably extract NLOS components. We will elaborate on the impact of the number of observations ($M$) later in this subsection.

\subsubsection{Symbol Detection} The key insight behind data detection in \acrshort{design} is that the overall received signal contains a strong contribution from the mainlobe with a stable amplitude and phase profile that does not significantly vary with sidelobe perturbations. In other words, the much weaker sidelobe signals bouncing off reflectors and reaching the receiver would have much lower power than the mainlobe signal.  \textcolor{black}{For instance, based on direct array factor calculation for a 16-element ULA with 4 randomly deactivated antennas (12 active), steered to broadside, the minimum amount of mainlobe-to-sidelobe-difference is 9.6 dB. This delta is much more pronounced when the extra path loss and reflection loss of NLOS components are taken into account. }

Hence, by taking into consideration Eq. (\ref{eq:received_nm}) and Eq. (\ref{eq:radiationP}) and assuming that the TX beam is aligned to the RX, we can write

\vspace{-3mm}
\begin{equation}
\begin{split}
y[n,m] & =  \bigg[\sqrt{N_{tx}^2 L^2 \alpha_0}+ \sum_{k=1}^{K} \sqrt{\alpha_k} R(m,\phi_k)\bigg] s[n,m]+ z[n,m] \\
& \cong N_{tx}L \sqrt{\alpha_0} s[n,m].
\end{split}
\label{eq:SNR}
\end{equation}
\vspace{-2mm}

Given the random perturbations in the sidelobe, the NLOS components will add up randomly at the receiver while the mainlobe direction provides a fixed and high directivity gain. Hence, the received constellation does not deviate too much from the nominal constellation realized when all received are ON. Fig.~\ref{fig:algo_sensing} illustrates the received constellation before and after applying compressive sidelobe patterns. We can observe that the received symbols have slightly lower power and random yet small phase fluctuations. Nevertheless, a scaled decision boundary can still be used for symbol detection. Since the antenna subset size ($L$) is known at the receiver, finding the scaled decision boundary is straightforward \textcolor{black}{by multiplying the original scaled boundary and $ L / N_{rx}$}. \textcolor{black}{We use a standard approach for channel estimation and symbol detection. Since the number of antennas OFF $L$ remains unchanged, the decision boundary for demodulation also stays the same. As a result, we perform data communication with the conventional method by using preambles at the start of each packet and apply channel equalization accordingly.
}.

We emphasize that the communication SNR can be modeled as SNR = $\epsilon_s \alpha_0 N_{tx}^2 L^2/\sigma^2$, \textcolor{black}{where $\epsilon_s$ is the average energy of symbols and $\sigma^2$ is the noise of Rx,} thus, the received SNR increases with the antenna subset size (i.e., the number of ON antennas). Hence there is a tradeoff between accurate symbol extraction and sensing performance: decreasing the antenna subset size hinders the communication SNR and may yield a higher symbol error rate but it will increase the random space for sidelobe sensing. We will explore it in Sec.~\ref{subsubsec:tradeoff}.

\subsubsection{Reflector Sensing Framework} \label{subsubsec:ref_sense_framework}
After detecting the data symbols, \acrshort{design} attempts to extract the AoA of the reflectors in the environment. Intuitively, such information is embedded in the small perturbation of the received signal vector around the constellation points. As depicted in Fig.~\ref{fig:algo_sensing}, such small perturbations are a function of compressive sidelobe in-use, the reflector angle, and noise. The impact of noise can be alleviated by averaging over several observations under a fixed compressive sidelobe pattern. Recall from Sec.~\ref{subsec:css} that we received multiple symbols under $m^\text{th}$ compressive sidelobe configuration. Therefore, as shown in Fig.~\ref{fig:algo_sensing}, the pattern of average power fluctuations under several different pseudorandom compressive beams acts as a fingerprint that hints at the underlying reflectors' characteristics. {Fig.~\ref{fig:algo_sensing} illustrates that such power variations are a unique signature for the reflector's angular location.

Next, we mathematically formulate \acrshort{design}'s AoA estimation scheme that builds on the insights explained above. Specifically, we can first find the amplitude of the signal captured by the mainlobe as follows:

\vspace{-2mm}
\begin{equation}
  \textcolor{black}{A_{ML}=\frac{1}{MN}\sum_m\sum_n \bigg|\frac{y[n,m]}{\tilde{s}[n,m]} \bigg| \cong N_{tx}L | \sqrt{\alpha_0} |},
    \label{eq:mainlobepower}
\end{equation}
\vspace{-2mm}

\noindent where $\tilde{s}[n,m]$ is the estimated symbol from the data extraction phase, $M$ is the total number of compressive sidelobe configurations and $N$ is the total number of measurements under each compressive beam. In other words, by taking the average of the measured signals under many different random sidelobe perturbations,  the result converges to the mainlobe signal, which is the unchanged portion of the transmission throughout. 
Further,  since the amplitude of the received signal is also a function of transmitted data (particularly in amplitude modulation schemes), we remove the impact of data symbols in calculating $A_{ML}$.

Next, we can isolate the impact of the NLOS components by removing the mainlobe contribution from the received signal: 

\vspace{-2mm}
\begin{equation}
    F [m]=\frac{1}{N}\sum_n{\bigg|\frac{y[n,m]}{\tilde{s}[n,m]} - A_{ML}\bigg|}.
    \label{eq:fingerprint}
\end{equation}
\vspace{-2mm}

Here  $F[m]$ is only dependent on NLOS paths and the sidelobe perturbations of the $m^\text{th}$ beam, \textcolor{black}{i.e., $F[m] \propto |\sum_{k=1}^{K} \sqrt{\alpha_k} R(m,\phi_k) |$}. Indeed, the vector $\mathbf{F}=[F[1], F[2], ..., F[M]]$ can be tested as the NLOS angle fingerprint explained above. Therefore, to find the angle of arrival of the most dominant NLOS path, we \textcolor{black}{leverage the principle of compressive sensing and} find the angle that offers the maximum correlation between the pattern of power perturbation vs compressive sidelobe beams in theory (i.e., $R(m, \phi)$ as defined in Eq.~(\ref{eq:radiationP})) against the measured perturbation fingerprint (i.e., $\mathbf{F}$):

\vspace{-2mm}
\begin{equation}
\textcolor{black}{ \phi_1^{*}={\rm argmax}_{\phi} \mathbf{F} \cdot |\mathbf{R}(\phi)|={\rm argmax}_{\phi} \sum_m F[m] \cdot |R(m, \phi)|},
\label{eq:argmax}
\end{equation}
\vspace{-2mm}

\noindent where $\phi_1^{*}$ is the estimated AoA of the dominant reflected path. If extracting $p$ NLOS paths is desired, then we extract the angle corresponding to the top $p$ peaks in \textcolor{black}{$\mathbf{F} \cdot |\mathbf{R}(\phi)|$}. It is expected that the estimation accuracy worsens for less pronounced paths as they are inherently weaker and leave a smaller footprint in sidelobe perturbations. \textcolor{black}{Fortunately, due to channel sparsity in mmWave bands, only a few paths may exist in practice. }

In \acrshort{design}, several parameters affect the NLOS extraction performance: First, increasing the number of measurements or observations (i.e., $N$) can help remove hardware non-idealities and channel noise, while improving sensing accuracy, consequently. Second, increasing the number of compressive sidelobes used during measurements (i.e., $M$) yields longer and more spatially diverse power perturbation signatures. Hence, \acrshort{design} can infer the AOA with finer granularity.  Third, increased radiation gain toward the sidelobe angles would definitely improve NLOS sensing. However, increasing the sidelobe power level would inevitably yield the theft of power from the mainlobe direction, hindering the underlying communication performance and reliability.

\textbf{Coherent vs Non-Coherent Sensing.} In principle, minute perturbations in power and phase can both be leveraged for NLOS sensing.  Extending the mathematical framework of \acrshort{design} to include complex signatures is straightforward. Specifically,  only slight changes to  Eq. (\ref{eq:mainlobepower}) to Eq. (\ref{eq:argmax}) are needed, i.e., we can write them as $A_{ML} = \frac{1}{MN} \sum_m \sum_n \frac{y[n,m]}{\tilde{s}[n,m]},$ $F [m]=\frac{1}{N}\sum_n (\frac{y[n,m]}{\tilde{s}[n,m]} - A_{ML}),$  and $\phi_1^{*}={\rm argmax}_{\phi} | \sum_m F[m] \cdot R^*(m, \phi) |$. }  However, for practical considerations, non-coherent sensing has certain advantages over its coherent counterpart: first, it is more resilient to the phase noise, carrier frequency offset, and synchronization  non-idealities commonly seen in commercial mmWave arrays ~\cite{rasekh2021phase} and second, it allows for aggregation of information across several packets that might not be tightly synchronized in phase.  We emphasize that if accurate phase measurements are available, \acrshort{design} can be easily extended to include phase readings in the AoA extraction framework, as discussed above.  \textcolor{black}{Inclusion of phase can pave the way for phase-based advanced sensing, like Doppler extraction and motion sensing}. Finally, while the above modeling considers a narrowband channel, the extension of \acrshort{design} to wideband channels with multiple subcarriers (i.e., OFDM) is straightforward. We will experimentally evaluate \acrshort{design} in both coherent and non-coherent cases  (under OFDM) in Sec.~\ref{subsec:cohe_wide_OFDM}. 


\begin{figure*}[h!]
     \begin{subfigure}{0.195\textwidth}
         \centering
         \includegraphics[width=\textwidth]{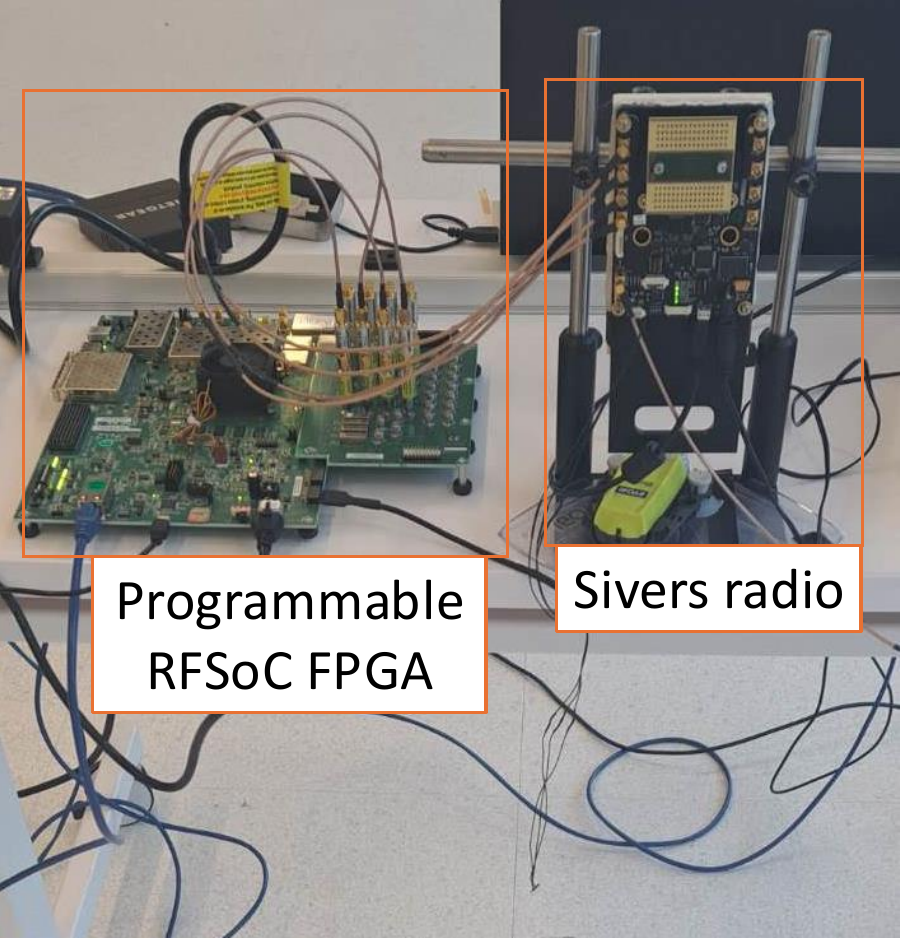}
         \caption{\textcolor{black}{Radio details.}}
         \label{fig:setup_array}
     \end{subfigure}
     \begin{subfigure}{0.39\textwidth}
         \centering
         \includegraphics[width=\textwidth]{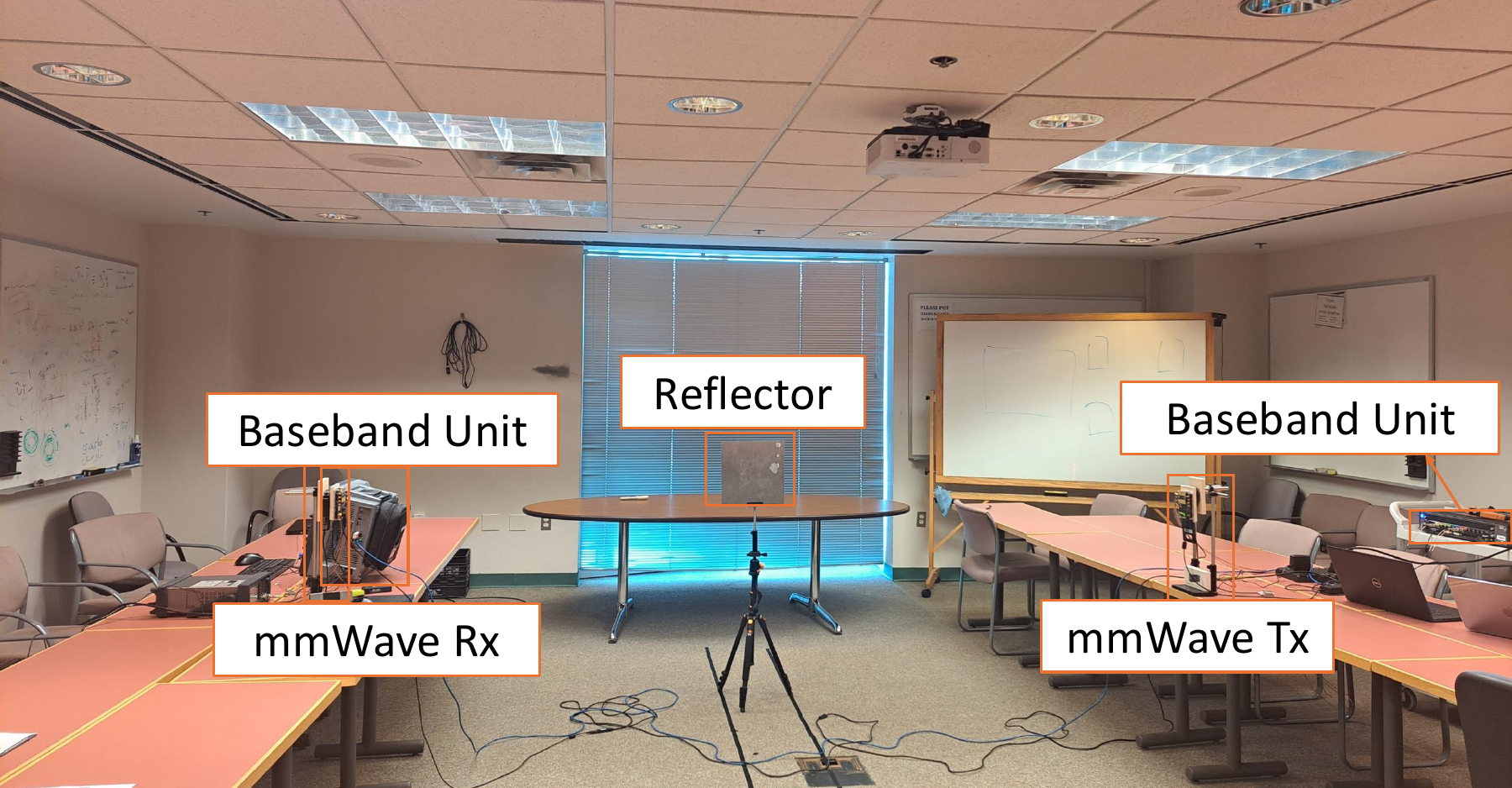}
         \caption{\textcolor{black}{Multipath-rich conference room}.}
         \label{fig:setup_conf_room}
     \end{subfigure}
     \begin{subfigure}{0.39\textwidth}
         \centering
         \includegraphics[width=\textwidth]{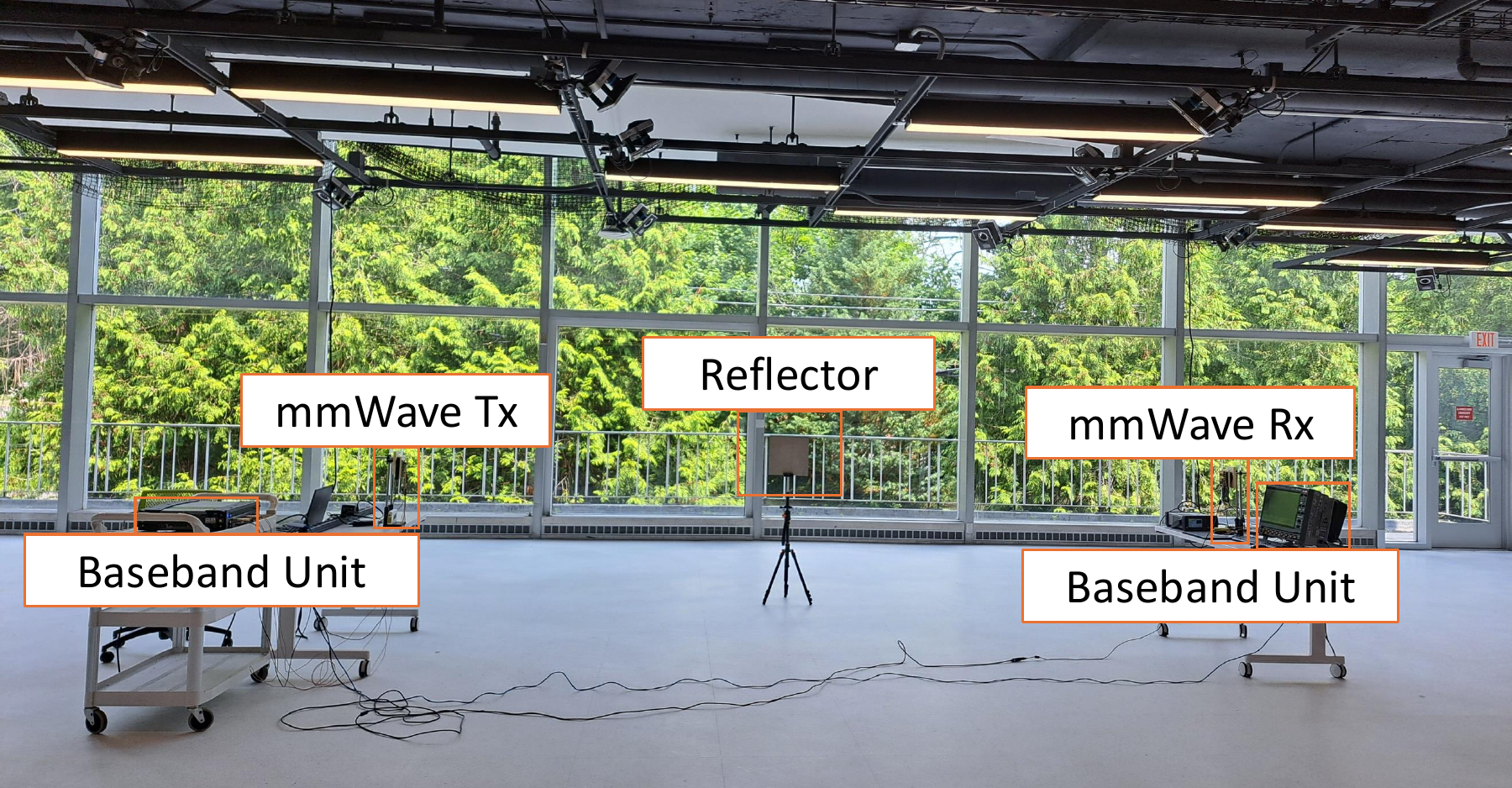}
         \caption{\textcolor{black}{Open lab space.}}
         \label{fig:setup_robo_lab}
     \end{subfigure}
                 \vspace{-1mm}
    \caption{Experimental setup with custom off-the-shelf mmWave arrays deployed in two different environments.}
                     \vspace{-3mm}
    \label{fig:setup}
\end{figure*}

\textbf{Overhead and Complexity of Sensing Framework.}
\acrshort{design} exploits received data symbols for sensing and hence does not incur any additional overhead in time. In terms of computational complexity, \acrshort{design}'s framework is lightweight, as also indicated from Eq. \eqref{eq:mainlobepower}-\eqref{eq:argmax}.  We note that $R(m, \phi)$ is a deterministic function of array geometry and can be known a priori. It is evident that the computational complexity scales linearly with the total number of sidelobe configurations. Furthermore, we argue that the number of compressive sidelobe patterns needed for accurate path sensing itself scales with $O({\rm log} N_{rx})$, where $N_{rx}$ is the number of antennas. Theoretically, with this number of patterns, the probability of error exceeding a certain range is very small~\cite{ramasamy2014compressive}. We emphasize that this complexity is much lower than the conventional path discovery with directional steering that scales with $O(N_{rx})$, thanks to the channel sparsity and properties of compressive sensing. 

Specifically, the idea of compressive sidelobe beams originates from compressive sensing in which several pseudorandom weight vectors are used for AoA extraction~\cite{rasekh2022design}. Specifically, in compressive sensing,  $w_{n} = e^{j \phi_{n}}$, where  $\phi_{n}$ is i.i.d. and follows a uniform distribution over $(0, 2\pi)$ to ensure $\mathbb{E}[w_{n}] = 0$, and  $Var[w_{n}] = 1$. This results in completely random radiation patterns that, unfortunately, fail to provide sufficient directivity gain for a communication link. Instead, in \acrshort{design}, by maintaining a high-gain mainlobe and perturbing the sidelobes, we achieve the best of two worlds: high-gain communication link from the mainlobe and compressive AoA estimation in sidelobe directions. Previous works rigorously proved that the path estimation task requires $O({\rm log}N)$ unique patterns~\cite{rasekh2022design}\textcolor{black}{, where $N$ is the array size}. Intuitively, by removing the impact of the mainlobe (see Eq.~\eqref{eq:fingerprint}), the residual pattern resembles the pseudorandom compressive beams from the literature; hence, the number of measurements/configurations scales with the same log factor. 
We skip the detailed proof due to space limitations.  \textcolor{black}{We note that a scanning sidelobe design—where a fixed mainlobe supports communication and a directional sidelobe sequentially scans across angles—incurs a higher overhead of $O(N)$, as it requires a separate beam configuration for each direction.}

This indicates that \acrshort{design} is scalable to future massive mmWave MIMO settings. 

\textcolor{black}{\textbf{Computational Complexity of Sensing Framework.}
The sensing algorithm is lightweight, with correlation being the most complex operation. If we note the number of theoretical compressive beams we use as $G$, the time complexity is $O(MGN_{rx})$, where $M$ is the number of compressive sidelobe configurations we use and $N_{rx}$ is the number of Rx antennas. The computational complexity mainly comes from theoretical compressive beam simulation. If we can simulate and store them beforehand, the running time complexity is only $O(MG)$. Therefore, it does not require extra computational resources and can be directly applied to current hardware. 
}


\subsubsection{Tradeoff between Communication and Sensing} \label{subsubsec:tradeoff}

In \acrshort{design}, the communication and sensing functions share the same front-end, time, frequency, and even the same waveform. Here, we elaborate on the tradeoffs between these two functionalities. In principle, the communication SNR depends on the directivity gain along the mainlobe direction, and the AoA resolution depends on the level of unique perturbations created along sidelobes angles. Given a fixed antenna array size, the mainlobe strength is directly related to the number of ON antennas ($L$). The random space for sidelobe perturbations, however, depends on $N_{rx} \choose L$ \textcolor{black}{, where $N_{rx}$ is the number of antennas in the RX array}. Therefore, an antenna subset size below $N_{rx}/2$ would not benefit communication or sensing. When $L$ increases ($L>N_{rx}/2$), the communication SNR improves but the random space for sensing shrinks, and vice versa. It is also evident that implementing \acrshort{design} on a mmWave node with a larger antenna array size (larger $N_{rx}$) is desired as it can create the same sidelobe random space while turning off fewer antennas, imposing less penalty on the communication SNR.  Hence, if there is an asymmetry between the array size at the two mmWave end nodes, running \acrshort{design} on the node that has a larger array is advantageous.  Interestingly, our evaluations suggest that randomly turning off only 2 antennas in a 64-element array creates sufficiently large random space for angular sensing of \textcolor{black}{1.5} degrees while imposing an SNR loss of less than 0.3 dB for the directional communication link.

\section{Implementation} \label{sec:implementation}

We implement \acrshort{design} on COTS mmWave hardware and conduct extensive over-the-air experiments at 60 GHz. 

\textbf{Hardware}. We use commercial mmWave phased arrays as front-ends at both TX and RX. Specifically, we use Sivers EVK06002 boards \cite{Sivers} that can operate in 57-71 GHz regime and are widely employed in previous mmWave research and development studies. These boards have $16 \time 4$ antennas for transmission and reception separately, as shown in Fig.\ref{fig:setup}(a). The array supports beam scanning in the azimuthal direction with the field of view of 90$\degree$, i.e., the mainlobe spans from -45$\degree$ to 45$\degree$ with a step size of about 1.5$\degree$. \textcolor{black}{The Sivers hardware has a maximum beam switching rate over 50 MHz from its datasheet. } We generate a compressive sidelobe beambook for any given mainlobe direction by randomly turning off a subset of antennas without modifying the phase shifters' values. \textcolor{black}{We randomly chose a random subset of antennas to configure each, without any selection optimization}. If not specified, we turn off four out of sixteen antennas, and we consider $63 \times 3 = 189$ different random subsets, i.e., sidelobe configurations. \textcolor{black}{Under each beam configuration, several symbols are captured. In other words, the beam switching rate is slower than the symbol rate.} We evaluate the impact of antenna subset size, number of sidelobe configurations,  and number of symbols per beam in Sec.~\ref{subsec:radio_exp}. 

\textbf{Experimental Setup.} We conduct experiments in two different environments. The first environment is a multipath-rich conference room with several tables, chairs, and whiteboards as shown in Fig. \ref{fig:setup}(b). The second environment is an open lab space, as shown in Fig. \ref{fig:setup}(c),  with a highly reflective glass wall and large square footage suitable for long-distance experiments. In both environments, we use a small metal reflector of size $0.3 \times 0.3$ m \textcolor{black}{(even smaller than a mini TV or a monitor)} and place them in various locations and orientations. We note that detecting this small reflector is more challenging compared with a large whiteboard or wall that was naturally present in those environments. 

\textcolor{black}{We have implemented  \acrshort{design} using both benchtop equipment as well as high-speed RFSoCs. In the former, we use an arbitrary waveform generator and real-time scope as baseband units on the TX and RX sides, respectively. In this case, the symbol rate is $25$ MHz and our frame structure contains 20 preamble symbols followed by 1500 data symbols. \textcolor{black}{The symbol rate is set to 25 MHz for one single tone. This is much higher than what standards today support. For mmWave bands, 5G NR commonly uses subcarrier spacings of 120 kHz and 240 kHz.} We also implemented \acrshort{design} programmable RFSoC FPGA as baseband units on both ends. The ADC sample rate is $3.52$ GHz, and the bandwidth is $1.1$ GHz with 192 active subcarriers. Our frame structure contains a standard IEEE 802.11 preamble and 64 data symbols.}

On the TX side, the mmWave array is connected to the baseband unit, sending modulated data streams with different modulation and coding rates. On the RX side, the downconverted IQ samples are captured and processed by another baseband unit. Differential IQ ports are connected for both TX and RX to remove the distortion and noise. The TX and RX clocks are not synchronized (the nodes are not wire-connected). \acrshort{design}'s algorithm is written in a single Python code controlling the Sivers array configuration and data capture. For all experiments, the center frequency is $60.48$ GHz. If not specified, we place the TX and RX 3 m apart (we evaluate the impact of TX-RX distance in Sec.~\ref{subsec:env_exp}). \textcolor{black}{We also use preambles as in a conventional communication system for packet detection.} The preambles are used for coarse carrier frequency offset removal, adopting conventional signal processing schemes from the literature~\cite{thomas2015blind}. The data symbols are used for joint data recovery and NLOS sensing. 


\section{Evaluation} \label{sec:eval}

\subsection{Compressive Sidelobe Radiation Patterns} \label{subsec:beam_gene}
First, we evaluate our compressive sidelobe generation technique by measuring the radiation pattern.

\textbf{Setup.} We place the transmitter and receiver at a distance of 1 m apart in an open space without any reflectors around (to resemble an anechoic chamber).  To measure the radiation pattern, we rotate the Rx from -50$\degree$ to 50$\degree$, which is slightly wider than the field of view of the phased array. 

\begin{figure}[t]
     \begin{subfigure}{0.24\textwidth}
         \centering
         \includegraphics[width=\textwidth]{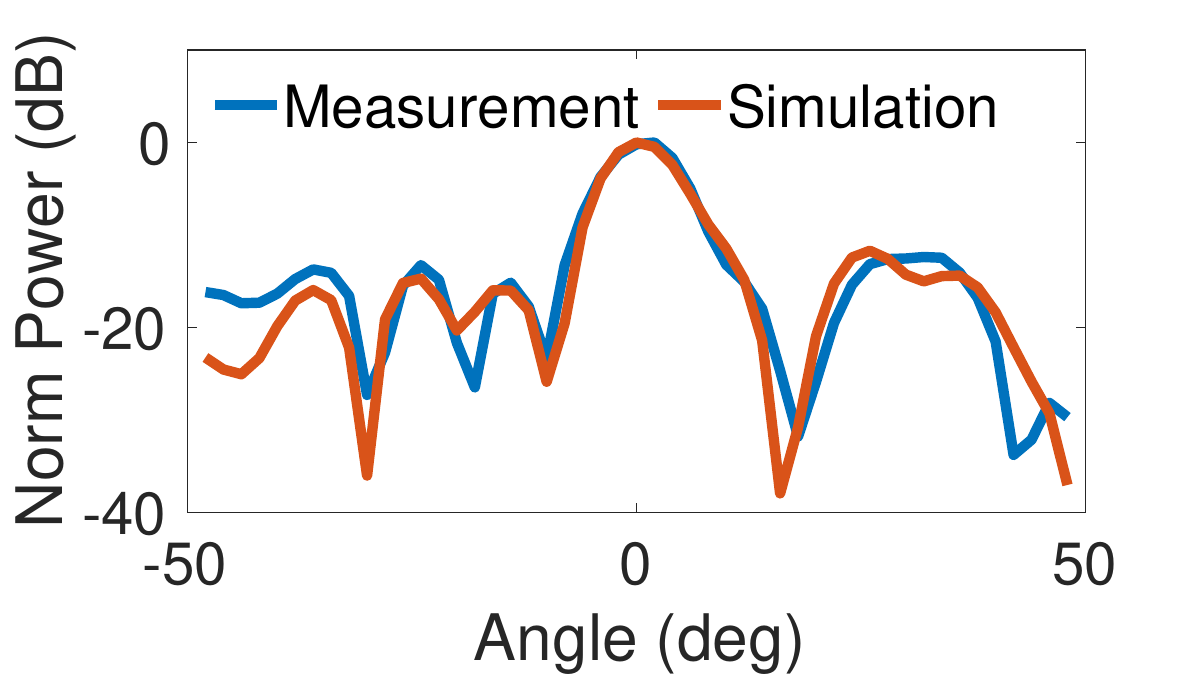}
                           \vspace{-6mm}
         \caption{}
         \label{fig:0_deg_beam}
     \end{subfigure}\hspace*{\fill}
     \begin{subfigure}{0.24\textwidth}
         \centering
         \includegraphics[width=\textwidth]{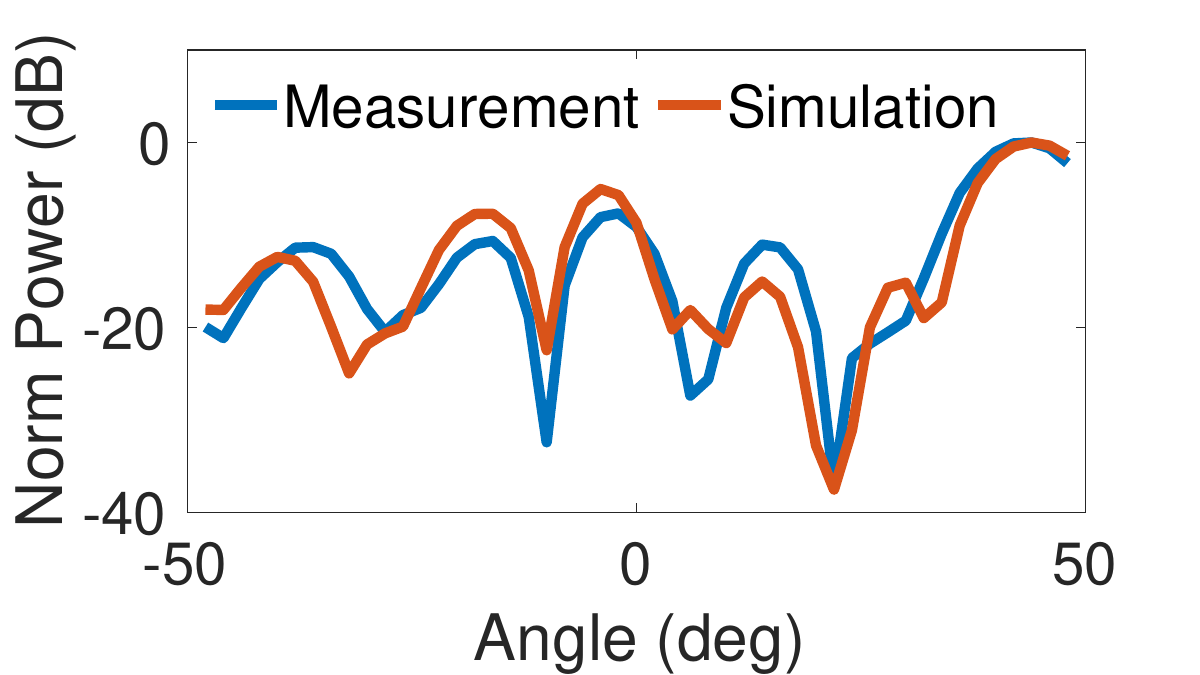}
                           \vspace{-6mm}
         \caption{}
         \label{fig:45_deg_beam}
     \end{subfigure}
     \begin{subfigure}{0.24\textwidth}
         \centering
         \includegraphics[width=\textwidth]{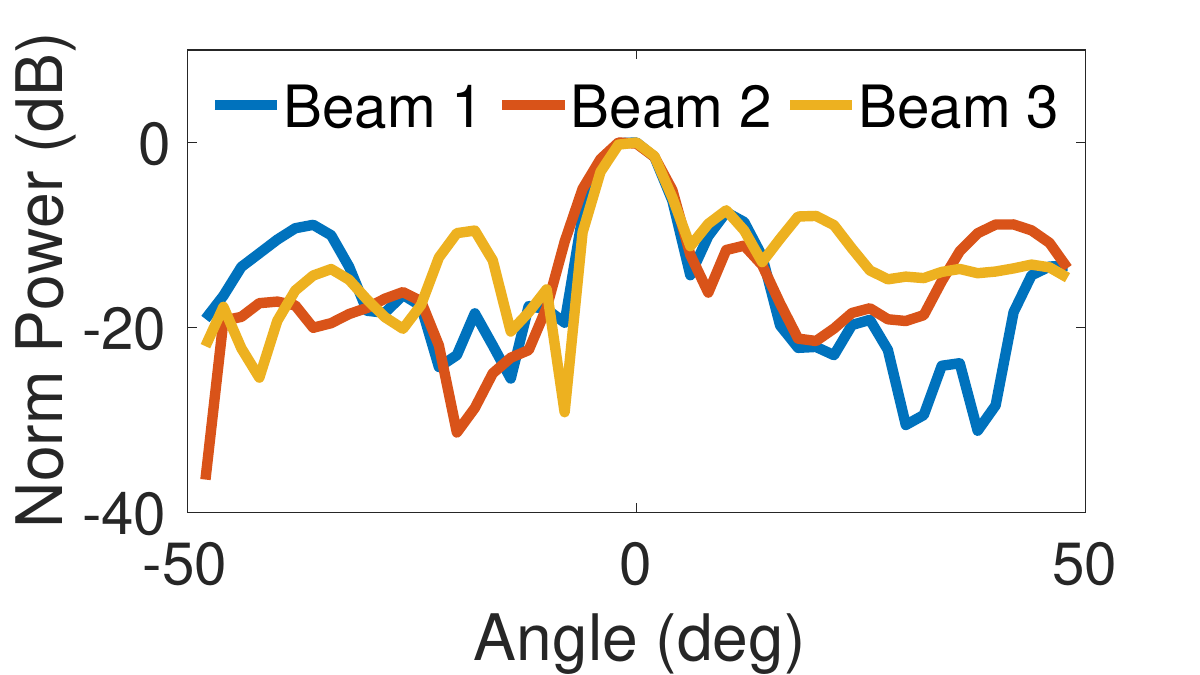}
         \vspace{-6mm} 
         \caption{}
         \label{fig:multi-side0}
     \end{subfigure}\hspace*{\fill}
     \begin{subfigure}{0.24\textwidth}
         \centering
         \includegraphics[width=\textwidth]{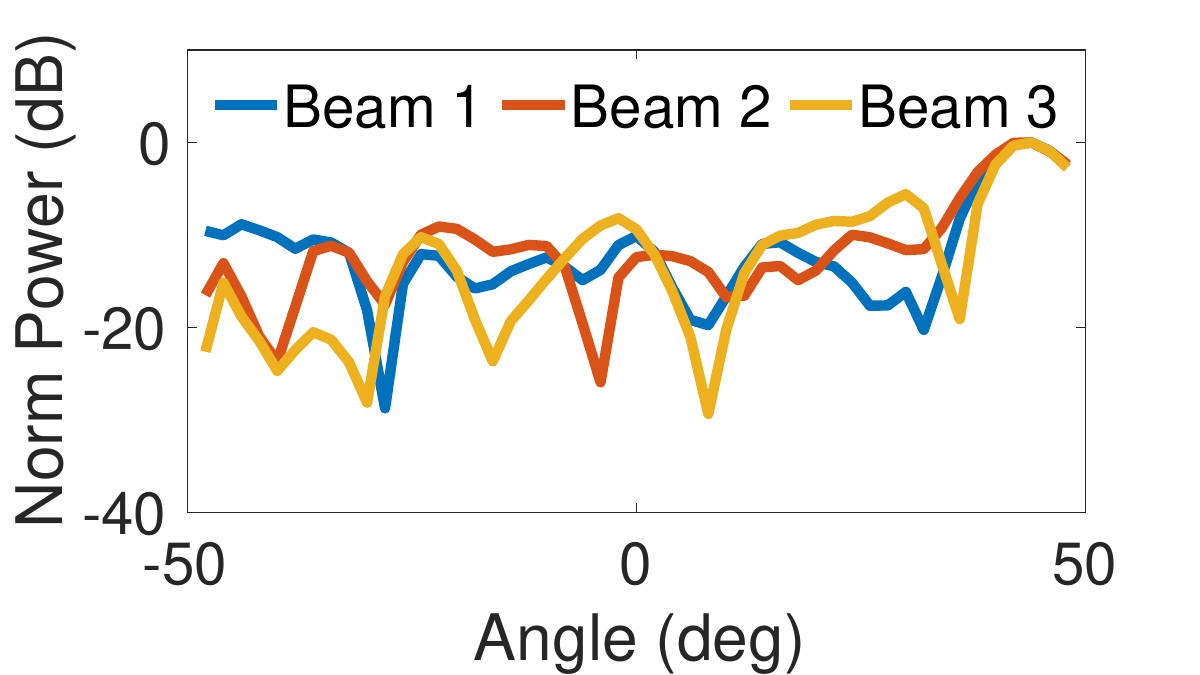}
          \vspace{-6mm} 
         \caption{}
         \label{fig:multi-side45}
     \end{subfigure}
          \vspace{-7mm}
     \caption{Measured and simulated compressive sidelobe beam patterns realized with random antenna subset switching; Mainlobe toward (a) 0$\degree$, (b) 45$\degree$; Sidelobe-varying beams with mainlobe toward (c) 0$\degree$, (d) 45$\degree$.}
     \vspace{-5mm}
    \label{fig:BeamPattern}
\end{figure}

\textbf{Results.} We measure the radiation pattern under different mainlobe directions and we randomly turn off 4 out of 16 antenna elements. The measured radiation patterns are shown in Fig.~\ref{fig:BeamPattern}. We show examples of two different mainlobe configurations and two distinct random sidelobe patterns under each mainlobe. In addition, we have simulated the theoretical radiation pattern using the knowledge of antenna geometry, inter-element spacing, and randomly chosen antenna subsets in measurements. We note that the internal non-idealities between antenna elements (e.g., due to variations in the lengths of the transmission lines in the circuit board connecting the RFIC to the antennas), require a one-time calibration for the array. We have performed antenna calibration based on the beambook provided by the manufacturer, \textcolor{black}{ i.e., by implementing a technique known as EiCal in the literature}~\cite{ano2024zero}. Fig.~\ref{fig:BeamPattern} illustrates good agreement between the simulated and measured radiation patterns. Note that we collected tens of different radiation patterns overall and validated the random sidelobe generation in all cases. Fig.~\ref{fig:multi-side0} and Fig.~\ref{fig:multi-side45} depict a few examples of measured sidelobe-varying beam patterns while the mainlobe remains unchanged toward 0$\degree$ and 45$\degree$, respectively. 

The results demonstrate a few key findings: First, the compressive sidelobe generation is able to maintain a directional dominant mainlobe. Second, strong sidelobe variations can be synthesized using a simple antenna ON/OFF switch. Third, the radiation pattern can be accurately predicted in simulations even with imperfect COTS arrays. We note that \acrshort{design}  relies solely on simulated compressive beams for AoA sensing as acquiring  measured radiation patterns of compressive beams is not realistic in practice. 

\subsection{\acrshort{design}'s JCS Performance} \label{subsec:main_exp}
Next, we evaluate the performance of \acrshort{design} in extracting the unknown symbols while angularly tracking the environment at the same time. 

\begin{figure}[t]
    \centering
    \begin{subfigure}{0.35\textwidth}
         \centering
         \includegraphics[width=\textwidth]{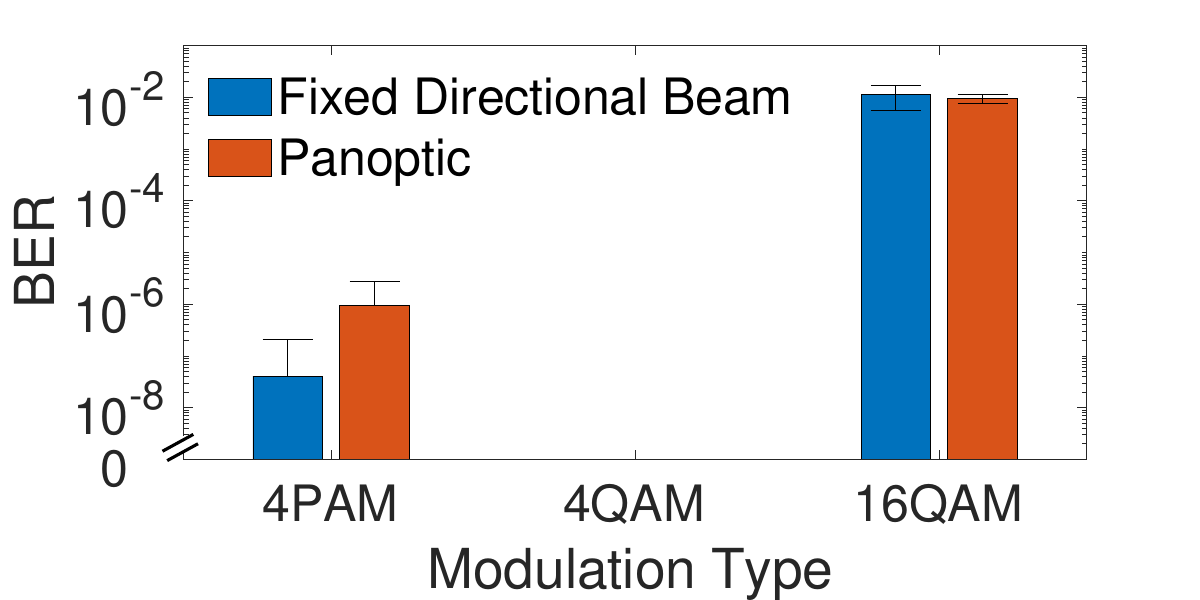}
                  \vspace{-6mm}
         \caption{BER performance}
         \label{fig:main_comm}
     \end{subfigure}
     \begin{subfigure}{0.35\textwidth}
         \centering
         \includegraphics[width=\textwidth]{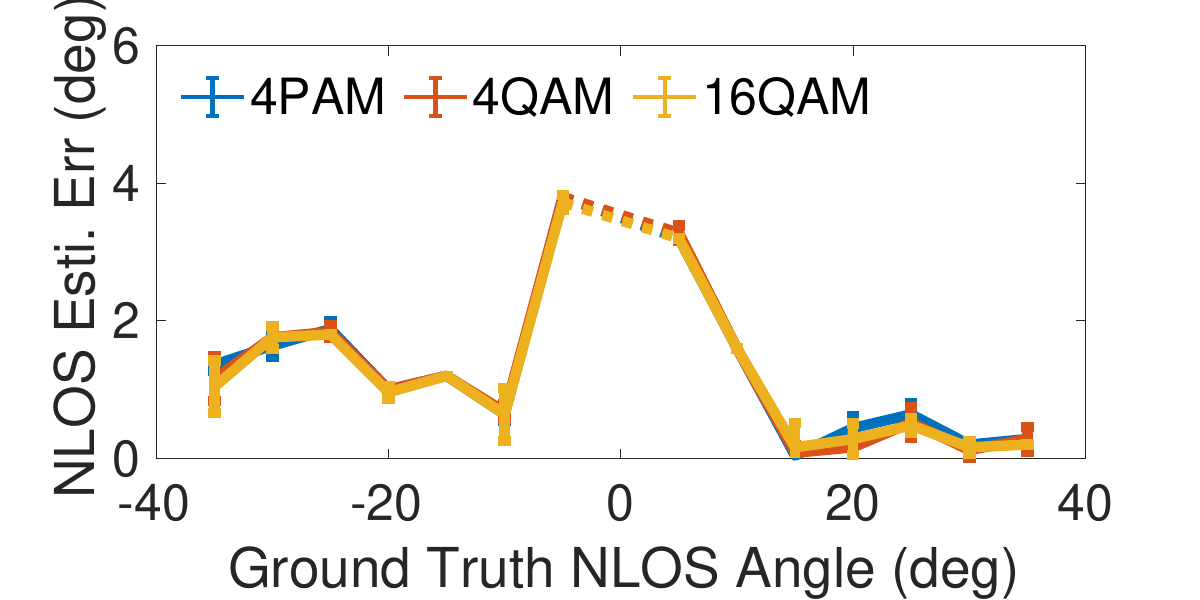}
                      \vspace{-6mm}
         \caption{Accurate NLOS AoA estimation}
         \label{fig:main_sense}
     \end{subfigure}
              \vspace{-1mm}
     \caption{\acrshort{design}'s JCS performance: (a) the measured BER under various modulations; (b) NLOS angle estimation accuracy.}
    \label{fig:main_comm_sense}
         \vspace{-6mm}
\end{figure}

\textbf{Setup.} We place the TX at the broadside of RX with a 3 m distance in a multipath-rich conference room (see Fig.~\ref{fig:setup}). In these experiments, \acrshort{design} attempts to angularly localize a reflector placed at different locations, i.e.,  from -40$\degree$ to 40$\degree$ (with 5$\degree$ step except at 0$\degree$ where the LOS path lies) relative to the receiver. At each reflector location, the transmitter sends modulated signals (with a total of $1.23 \times 10^8$ bits) and the receiver captures the signal with 189 compressed sidelobes beams. Each compressed beam is generated by randomly turning off $4$ out of 16 antennas (i.e., $L=12$). 

\textbf{Result.} Fig. \ref{fig:main_comm} summarized the BER performance under different modulation types. For comparison purposes, we repeat the experiments with a conventional directional beam (all antennas are ON creating a strong mainlobe without any sidelobe perturbations) that supports communication only. We observe that \acrshort{design} yields a slight increase in BER under 4PAM but the BER remains unchanged with other modulations. In particular, for 4PAM, the mean BER for directional beam and compressive sidelobe beam are 4.0967$\times10^{-8}$ and 9.4954$\times10^{-7}$, respectively. For 4QAM, the mean BER for the directional beam and compressive sidelobe beam is 0. \textcolor{black}{The better performance of 4QAM compared with 4PAM is owed to the carrier frequency offset mitigation algorithm implemented in our setup~\cite{thomas2015blind}.} As expected, the BER is higher for 16QAM. Interestingly, spatial diversity offered by sidelobe perturbations in \acrshort{design} even slightly improves the BER results. Hence, in summary, a compressive sidelobe beam achieves similar BER performance compared with adopting conventional fixed directional beams, regardless of modulation type.

However, unlike conventional fixed beams, \acrshort{design} leverage random sidelobe perturbation for NLOS sensing. The estimation accuracy of NLOS AoA is shown in Fig.\ref{fig:main_sense}. The mean estimation error for all three modulations is below 2$\degree$ for all ground truth NLOS angles except for $\pm 5\degree$. The higher estimation error at $\pm 5\degree$ is rooted in the fact that at those angles, the NLOS path is close to the LOS angle (which is at $0\degree$). In general, the NLOS sensing results are promising. 
The mean angle error for 4PAM, 4QAM, and 16QAM is 1.28$\degree$, 1.25$\degree$, and 1.23$\degree$, respectively. Moreover, the standard deviations for 4PAM, 4QAM, and 16QAM are 1.08$\degree$, 1.12$\degree$, and 1.10$\degree$, respectively. Hence, the NLOS sensing performance is robust to various phase/amp modulations. 

The NLOS sensing in \acrshort{design} offers a reliable backup path for the mmWave communication link without incurring any additional overhead. Indeed, if the LOS path is blocked, the mmWave nodes can automatically adapt their beams toward the known NLOS path. We highlight that with our COTS 16-antennas array, the half-power beamwidth is 6.6$\degree$, and the NLOS AoA estimation error of 2$\degree$ would result in about $1.02$ dB loss in maximum achievable directivity gain. In summary, \acrshort{design} achieves high reflector sensing accuracy with a negligible loss on the accurate data detection performance. 
\subsection{Impact of \acrshort{design}'s Design Parameters} \label{subsec:radio_exp}

In Sec.~\ref{sec:design}, we illustrated that a few key parameters affect \acrshort{design}'s data reception and NLOS sensing capabilities. Namely, the number of compressive sidelobe beams (M), the antenna subset size (L), and the number of data symbols captured under any given beam (N). Here, we experimentally evaluate the impact of these parameters. To this end, we use the setup explained in Sec.~\ref{subsec:main_exp} and use 4QAM for modulation (the same trends and findings hold for other modulations).

\begin{figure}[t!]
     \medskip
     \begin{subfigure}{0.24\textwidth}
         \centering
         \includegraphics[width=\textwidth]{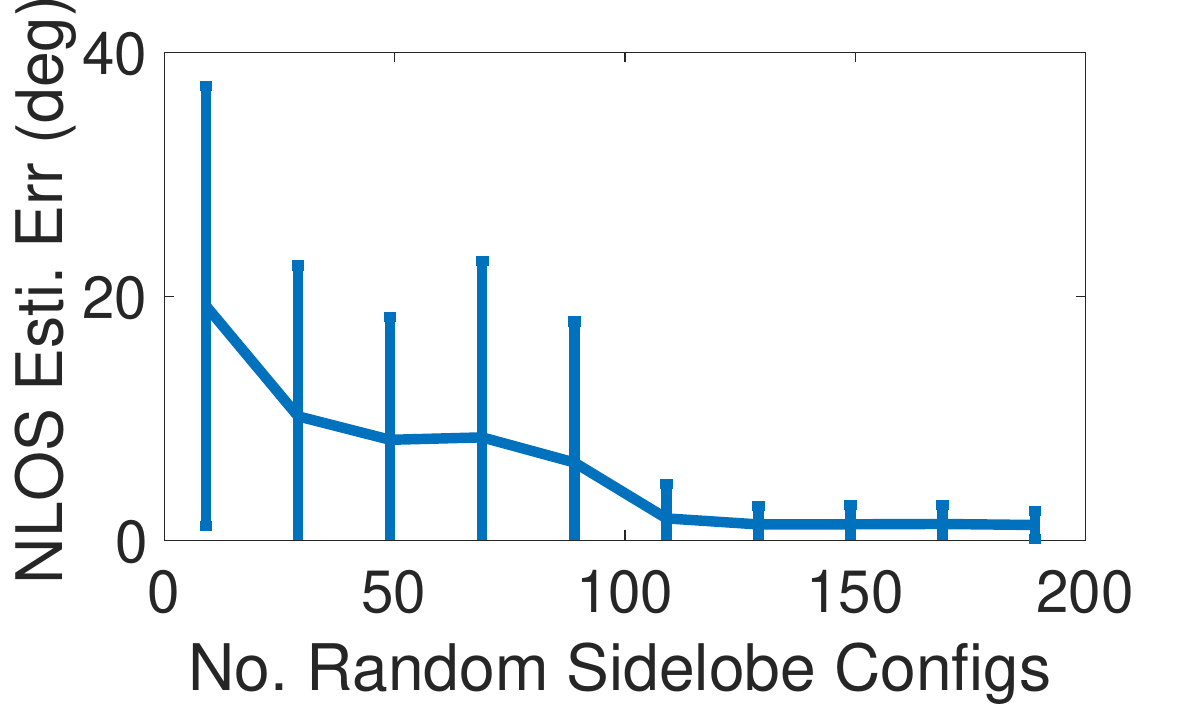}
                  \vspace{-6mm}
         \caption{}
         \label{fig:micro_BeamNum}
     \end{subfigure}\hspace*{\fill}
     \begin{subfigure}{0.24\textwidth}
         \centering
         \includegraphics[width=\textwidth]{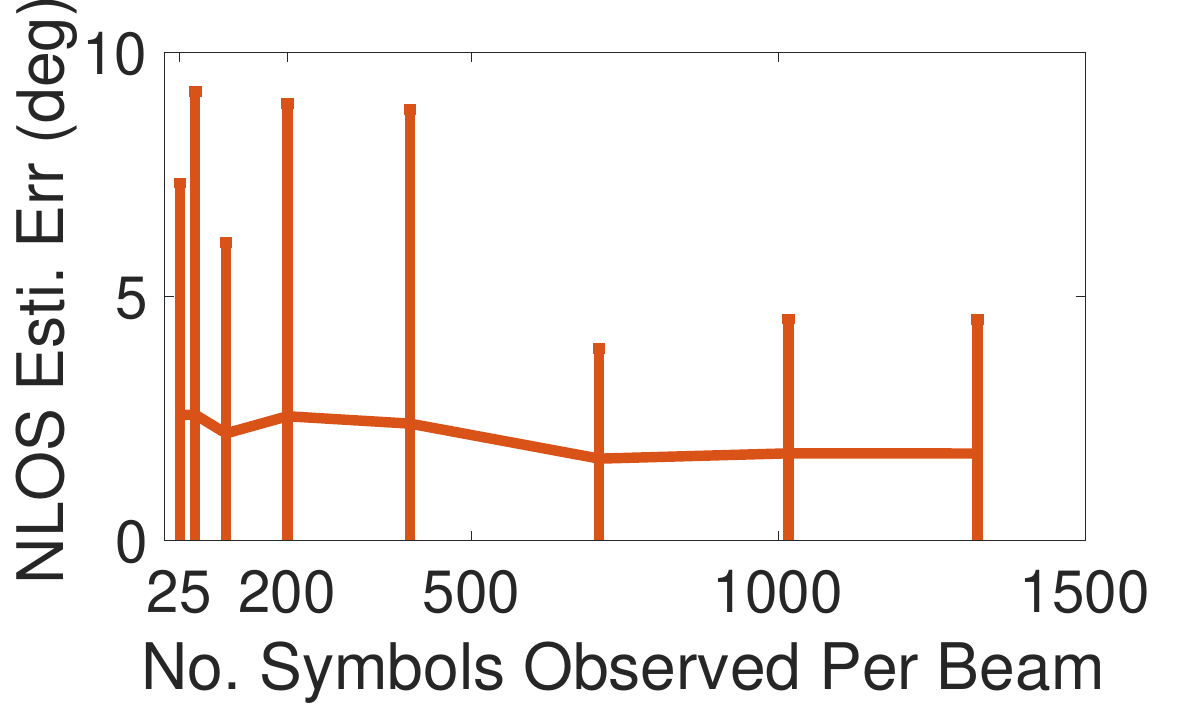}
                  \vspace{-6mm}
         \caption{}
         \label{fig:micro_SymNum}
     \end{subfigure}
     \vspace{-5mm}
    \caption{Impact of \acrshort{design} parameters on NLOS estimation error: (a) Number of random sidelobe configuration tested; (b) Number of data symbols (measurements) captured per beam.}
    \label{fig:micro_nub_control_large_time}
    \vspace{-4mm}

\end{figure}

\subsubsection{Number of random sidelobe configurations ($M$)}
First, Fig.\ref{fig:micro_BeamNum} depicts the NLOS angle estimation accuracy with the varying number of randomly generated sidelobe configurations. We observe that the estimation error decreases monotonically with more sidelobe configurations, which matches well with the theoretical analysis in Sec.\ref{subsubsec:ref_sense_framework}. Further, with 100+ configurations, the error reaches a plateau. Hence, beyond this point, there is only negligible improvement in NLOS extraction when adopting more beam patterns, but rather, this would yield a higher sensing and detection delay. In practice, \acrshort{design} adapts the number of sidelobe configurations based on the desired sensing resolution and the scale of mobility in the environment.

\subsubsection{Number of data symbols per beam ($N$)}
Second, Fig.\ref{fig:micro_SymNum} shows the impact of the number of data symbols captured per compressive beam at the receiver. \textcolor{black}{We note that the symbol rate is always higher than the antenna switching rate in Panoptic, and hence, several symbols can be captured under a given beam configuration.Here we set the number of beams to 189.} Recall from Eq.~\eqref{eq:fingerprint} that several measurements with a fixed beam help with averaging out noise and better mapping the sidelobe perturbations to the NLOS path signature. This is also reflected in our experimental results in Fig.\ref{fig:micro_SymNum}. We observe that the mean sensing error decreases with more observations per beam. With 25 symbols per beam, the mean NLOS angle error is 2.57$\degree$, very close to the final value. The average reflector estimation error decreases below 2$\degree$ after the symbol number exceeds 700. Also, the standard deviation decreases with more observations, which indicates a robust estimation. \textcolor{black}{In Panoptic, the symbol rate does not play any major role in the sensing accuracy. Depending on the symbol rate for the communication needs, we can adapt the antenna switching rate accordingly to allow for capturing the desired number of symbols in one beam configuration.}

These results indicate \acrshort{design}'s ability to support beam switching at the sub-packet level, avoiding the untimely update problem caused by long data packets and systems using only known preamble. In other words, \acrshort{design} achieves accurate NLOS sensing independent of the packet length, the presence or lack of known preambles, and other protocol-specific parameters. Instead, \acrshort{design} obtains rich information about the medium by analyzing the captured data symbols themselves at a much faster sampling rate.

\subsubsection{Antenna subset Size ($L$)}
Finally, we evaluate the impact of antenna subset size, i.e., the number of randomly selected OFF antennas. To this end, we repeat the experiments by turning off 1,2,3, and 4 antennas (out of 16) while placing a reflector at various angular locations within the range of -30 to 30 degrees (relative to TX). Recall that when all antennas are ON, no sidelobe perturbation happens, and hence NLOS sensing is not possible. We note that the total number of possible sidelobe configurations is only $M$=${16} \choose{1}$ =16 and $M$=${16} \choose{2}$ =120, for the case with 1 and 2 antennas OFF, respectively. For a larger number of OFF antennas, the random beam space size is much larger than the beambook memory in the Sivers arrays, and hence we limit the number of tested beam configurations to $M=3\times63=189$. 

 As Fig.\ref{fig:micro_ant_off_sensing} shows the reflector angle estimation error decreases as the number of OFF antennas increases. This matches well with our mathematical foundation in Sec.\ref{subsubsec:tradeoff} that the more number of antennas off, the larger the number of possible beam patterns, and the higher the sensing accuracy. The mean angle error is 1.97$\degree$ with 3 antennas OFF and 0.89$\degree$ with 4 antennas OFF. We note that the lower average error compared with results in Sec.~\ref{subsec:main_exp} is due to the fact that the considered angular range for the reflector is smaller in this experiment. 

\begin{figure}[t!]
     \medskip
     \begin{subfigure}{0.24\textwidth}
         \centering
         \includegraphics[width=\textwidth]{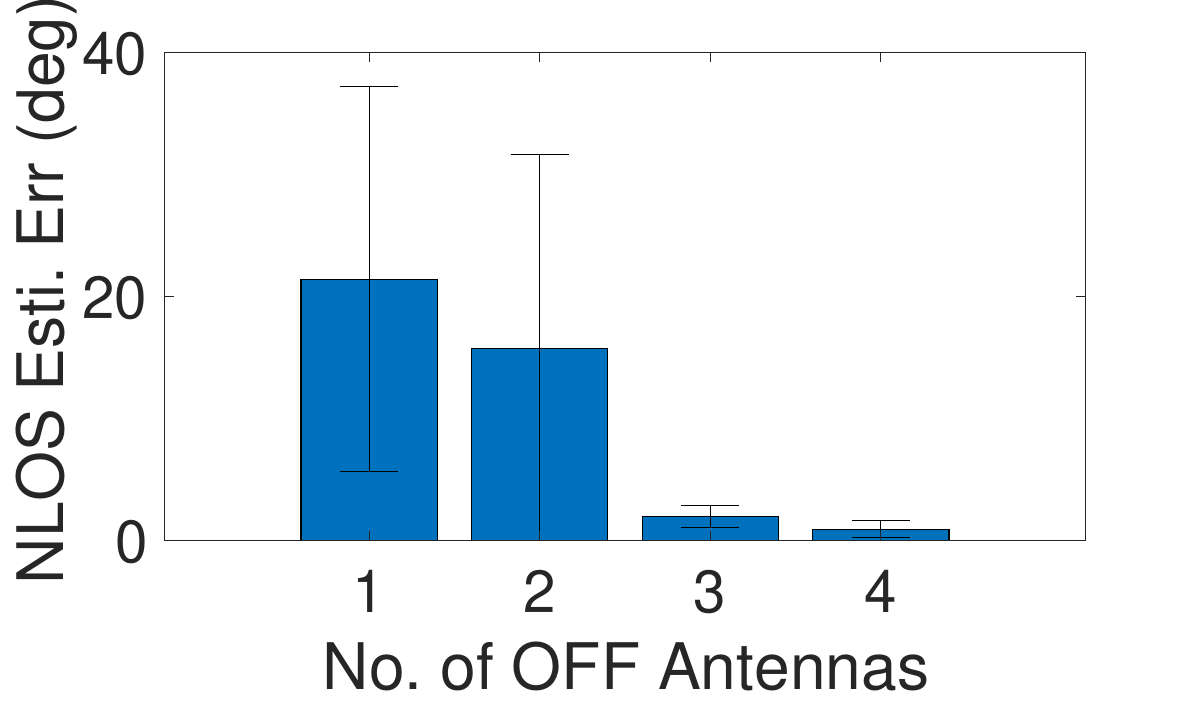}
                  \vspace{-6mm}
         \caption{}
         \label{fig:micro_ant_off_sensing}
     \end{subfigure}\hspace*{\fill}
     \begin{subfigure}{0.24\textwidth}
         \centering
         \includegraphics[width=\textwidth]{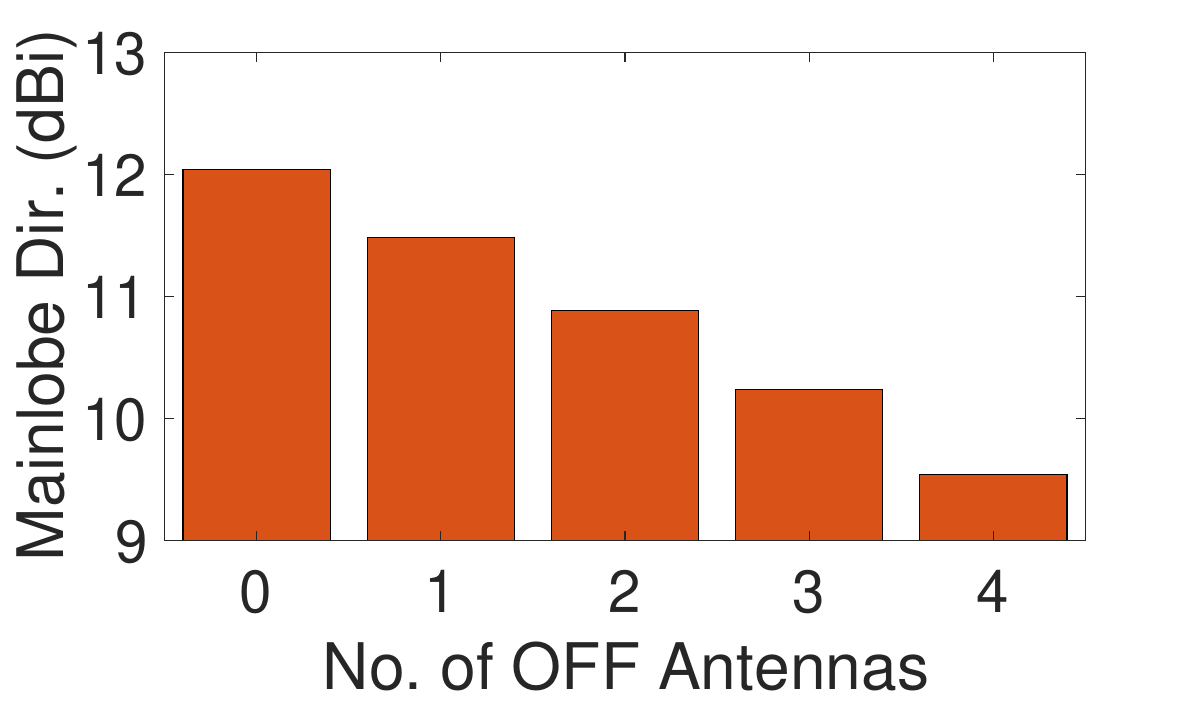}
                  \vspace{-6mm}

         \caption{}
         \label{fig:micro_ant_off_comm}
     \end{subfigure}
     \vspace{-6mm}
    \caption{Impact of random antenna subset size: turning off more antennas (a)  yields a larger random beamspace and better NLOS angular localization; (b) decreases the communication (mainlobe) directivity gain.}
     \vspace{-5mm}

    \label{fig:micro_nub_control_antenna_off_num}
\end{figure}

\begin{figure}[t!]
     \medskip
     \centering
     \includegraphics[width=0.36\textwidth]{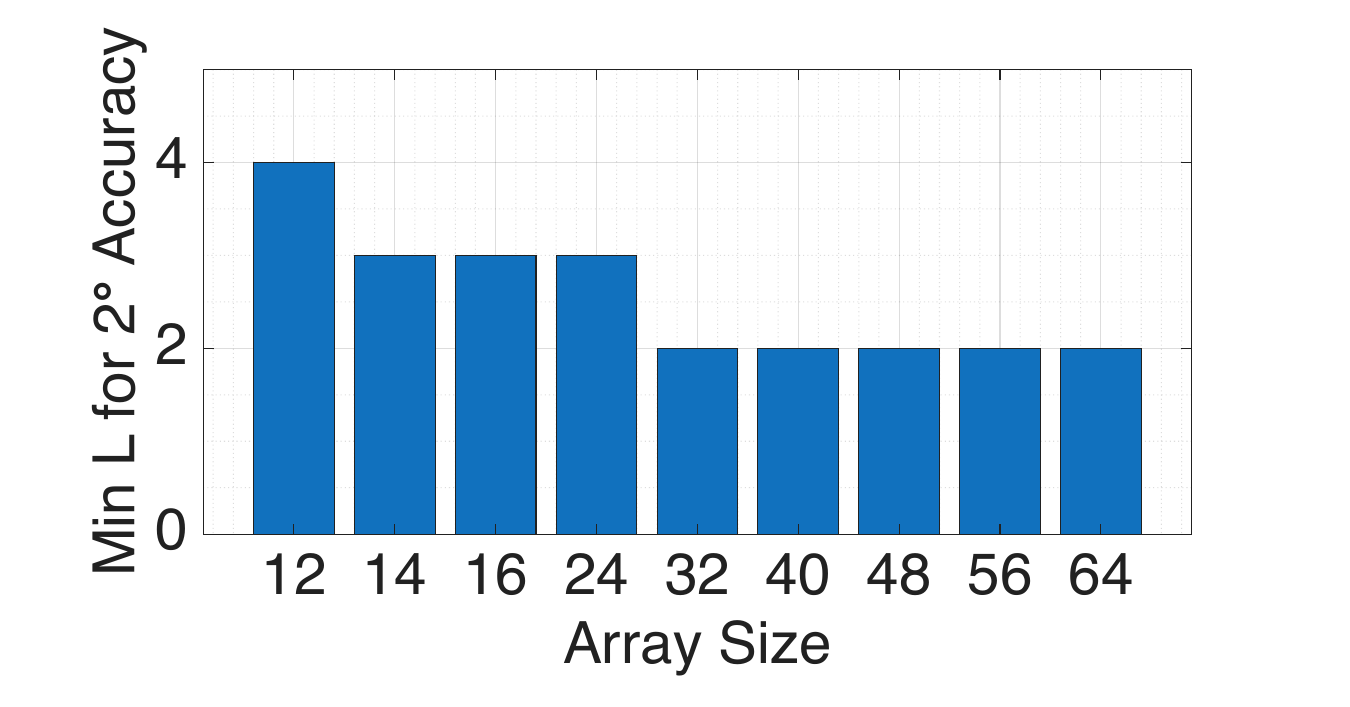}
    \vspace{-1mm}
    \caption{\textcolor{black}{Scaling to large arrays: as the array size grows, one can realize the desired sensing accuracy by turning off fewer antennas.}}
     \vspace{-5mm}
      \label{fig:micro_ant_off_vs_ant_total}

\end{figure}

On the other side, when turning off more antennas, the mainlobe directivity gain is reduced, which is also captured in Fig.\ref{fig:micro_ant_off_comm} \textcolor{black}{(all antennas ON in the leftmost)}. This may hinder the data detection performance with \acrshort{design}. In this particular experiment, the BER remained at 0 (out of $10^8$ transmitted bits) in all cases as the gain of the mainlobe was sufficiently high to close a 3-meter link between the TX and RX even when using 12 antennas only. We evaluate the impact of TX-RX distance in Sec. \ref{subsec:env_exp}. \textcolor{black}{We underline that the communication penalty becomes negligible with a larger antenna array because fewer antennas have to be turned OFF. For example, the randomness space size for the 16-element array with 4 antennas OFF is $16 \choose {4}$$=1820$, and the size for a 64-element array with 2 antennas OFF is $64 \choose {2}$$=2016$. However, turning off 2 out of 64  antennas decreases the mainlobe directivity (and consequently the SNR of the data link) roughly by  $20 \log_{10}{\frac{62}{64}} = -0.2758$ dB. According to the MCS -EVM table in IEEE 802.11ad MCS~\cite{80211ad}, such an SNR reduction will not trigger an MCS downgrade. }

\textcolor{black}{Next, we look at the minimum antenna subset size ($L$) required to achieve an average angle error of less than 2$\degree$. Our simulation considers a Tx-Rx distance of 2 m using 189 random beams. 200 randomly located reflector configurations are considered, such that they are all within +/- 45 degrees of the TX/RX nodes and the reflectors create a specular NLOS path between the TX and RX.  We repeat this simulation for different array sizes from 12 to 64.  The results are shown in Fig. \ref{fig:micro_ant_off_vs_ant_total}. We observe that the number of antennas OFF required for maintaining the same sensing accuracy slightly decreases with a larger array size. For example, a 12-element array requires 4 antennas to be turned off, but a 64-element array only needs 2. We also note that the antenna array with 8 or 10 antennas fails to achieve the sensing resolution even when turning off half of its elements (maximizing the randomness space). The best achievable angle estimation errors are 4.12° and 2.69°, respectively. In practice, the antenna subset size can be adapted in \acrshort{design} to balance between angular localization accuracy and the desired BER.}


\subsection{Microbenchmarks} \label{subsec:env_exp}

\subsubsection{Impact of Tx-Rx Distance}
In all previous experiments, the TX-RX distance was set to 3 m. Here, we evaluate how \acrshort{design}'s performance extends to longer-range communication and sensing. To this end, we repeat our experiments in an open space (see Fig.~\ref{fig:setup}), to push the range up to 15 m. We keep the radio power gains fixed throughout the experiment.

Fig. \ref{fig:micro_Distance_comm} depicts the result. We observe that as expected, the BER worsens with larger distances. However, interestingly, the BER results are comparable when applying a fixed directional beam (maximum mainlobe directivity) vs. when adopting \acrshort{design} that steals power from the mainlobe direction for the sake of sidelobe modulation. Specifically, in distances below 7.5 m, the BER is 0 for both the directional and compressive sidelobe beams. At a 10 m distance, the BER for directional and compressive sidelobe beam is around 4.5$\times10^{-4}$. When the distance is 12.5 m or larger, the BER gap between the directional beam and compressive sidelobe beam widens as in such low SNR regimes an additional mainlobe directivity can make a more significant difference.

\begin{figure}[t!]
     \medskip
     \begin{subfigure}{0.24\textwidth}
         \centering
         \includegraphics[width=\textwidth]{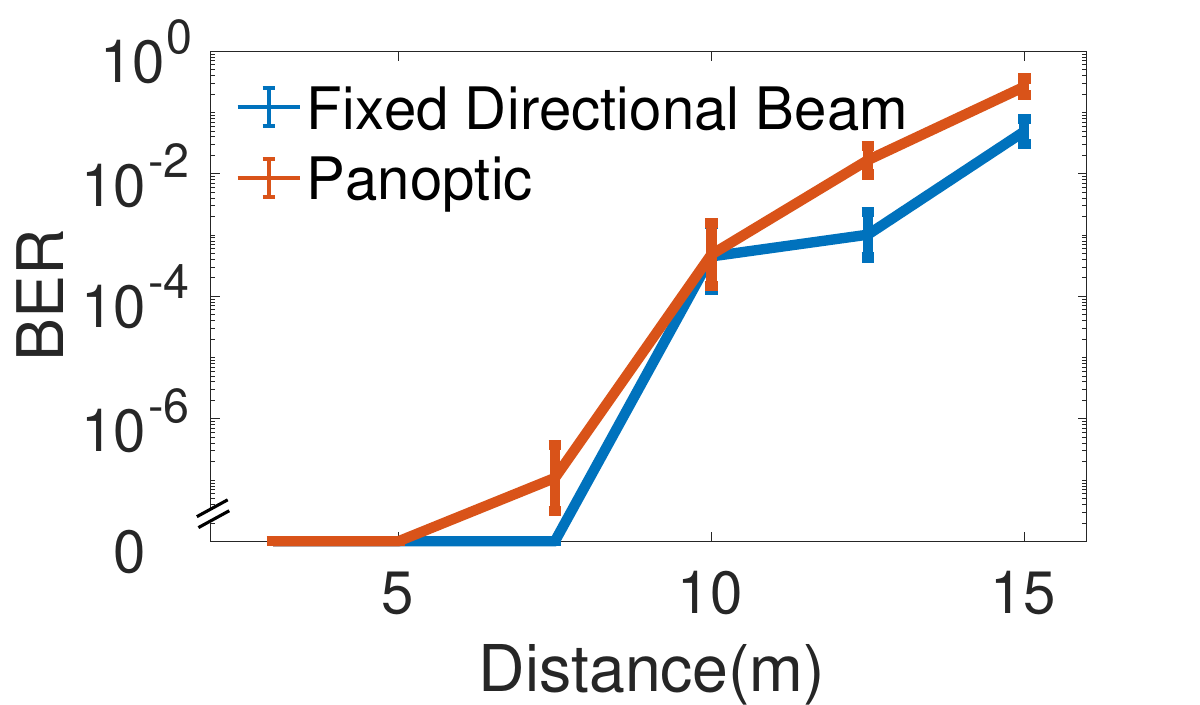}
                  \vspace{-6mm}
         \caption{Data Reception}
         \label{fig:micro_Distance_comm}
     \end{subfigure}\hspace*{\fill}
     \begin{subfigure}{0.24\textwidth}
         \centering
         \includegraphics[width=\textwidth]{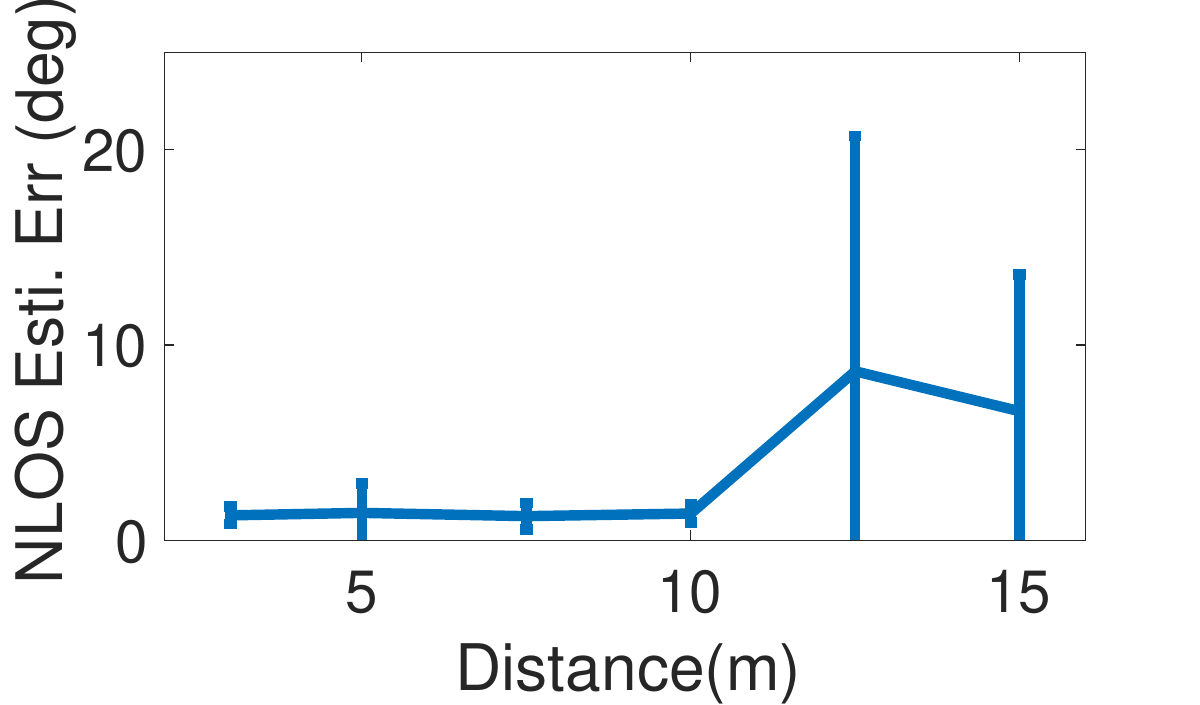}
                          \vspace{-6mm}
         \caption{NLOS Sensing}
         \label{fig:micro_Distance_sensing}
     \end{subfigure}
          \vspace{-6mm}
    \caption{Impact of TX-RX range on \acrshort{design}.}
     \vspace{-5mm}
    \label{fig:micro_distance}
\end{figure}

Fig. \ref{fig:micro_Distance_sensing} shows the NLOS sensing accuracy degrades beyond 10 m ranges. The mean angle estimation error is 1.34$\degree$ at 10m but it increases rapidly to 8.65$\degree$ at 12.5 m. The reason is that errors in symbol detection would inevitably lead to inaccurate reflector fingerprint extraction and NLOS angular localization, consequently. Indeed, as we explained in Sec. \ref{subsubsec:ref_sense_framework}, a successful symbol detection is the foundation for accurate NLOS angle estimation.

To summarize, \acrshort{design}'s communication and sensing performance remains promising in long ranges. Ultimately, in very low SNR regimes, the data perception capability is hindered and its corresponding errors negatively affect the ability of \acrshort{design} to accurately extract the NLOS components. We emphasize that with automatic gain control at the mmWave nodes, distances beyond 10m are achievable too. In these experiments, we kept the amplifier gain fixed to better understand the underlying distance-related trends. Nevertheless, our setup supports an additional $45$ dB of amplification gain that, if used, can extend the range significantly.

\subsubsection{Impact of TX-RX Relative Angular Location} 
So far, we placed TX and RX at the broadside of each other and varied the reflector location. Here, we change the TX-RX relative angular location as depicted in Fig. \ref{fig:TRxMobility_scene} and investigate the performance of \acrshort{design}. At each LOS configuration, we try several reflector placements. From Fig. \ref{fig:TRxMobility_result}, we observe that the NLOS AoA estimation accuracy is robust to such nodal configurations and the mean estimation error remains less than 2$\degree$ regardless of TX-RX angular location. Further, the BER remains all 0 for 4QAM (not shown).

\begin{figure}[t]
     \medskip
     \begin{subfigure}{0.24\textwidth}
         \centering
         \includegraphics[width=\textwidth]{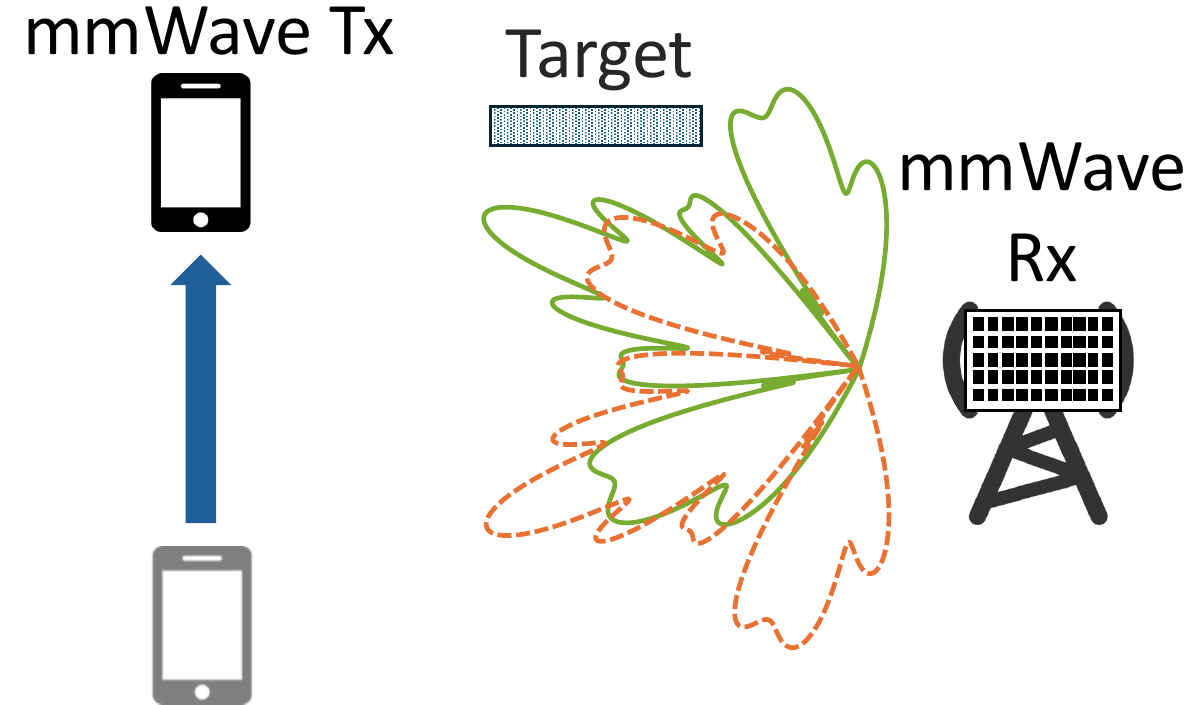}                  \vspace{-6mm}

         \caption{}
         \label{fig:TRxMobility_scene}
     \end{subfigure}\hspace*{\fill}
     \begin{subfigure}{0.24\textwidth}
         \centering
         \includegraphics[width=\textwidth]{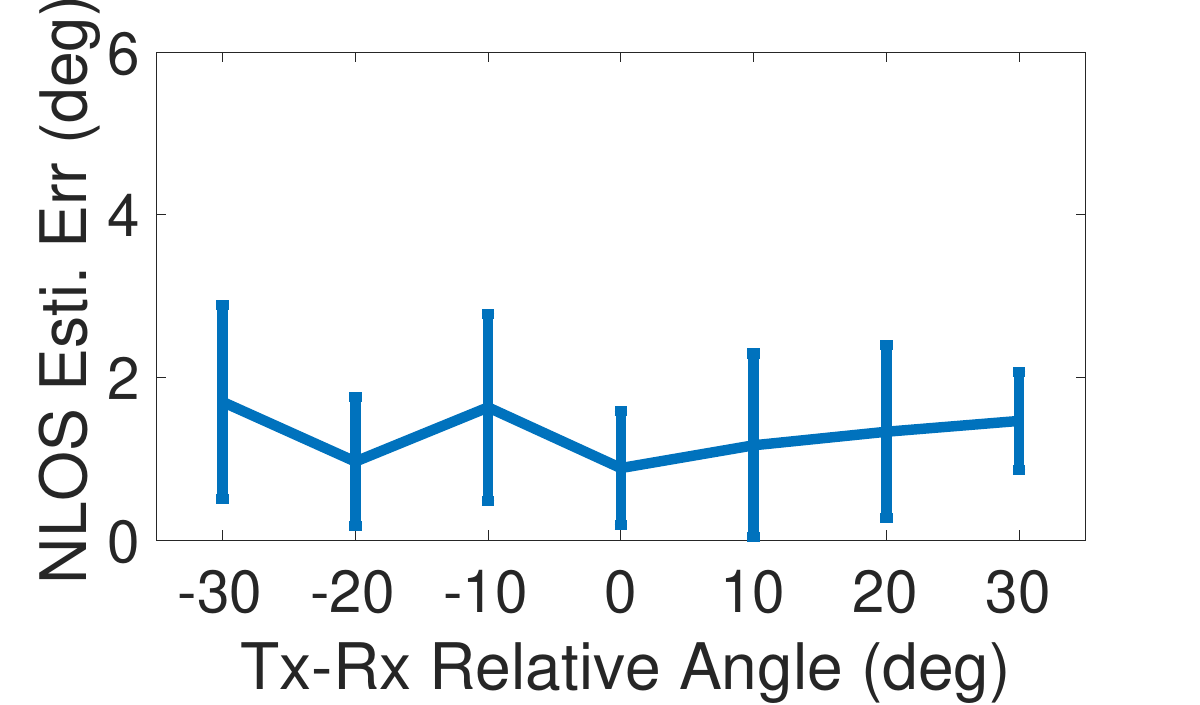}
                  \vspace{-6mm}

         \caption{}
         \label{fig:TRxMobility_result}
     \end{subfigure}
             \vspace{-6mm}
    \caption{\textcolor{black}{The impact of Tx-Rx relative angle: (a) Schematic of the setup; (b) NLOS angle estimation error.}}
    \vspace{-4mm}
    \label{fig:TRxMobility}
\end{figure}

\subsubsection{Impact of Nodal Mobility} 
In all previous experiments, the transmitter was fixed while the environment (reflector) kept moving. Here, we ask the question: What happens if both the environment and the transmitter are mobile? 

\begin{figure}[t]
     \medskip
     \begin{subfigure}{0.24\textwidth}
         \centering
         \includegraphics[width=\textwidth]{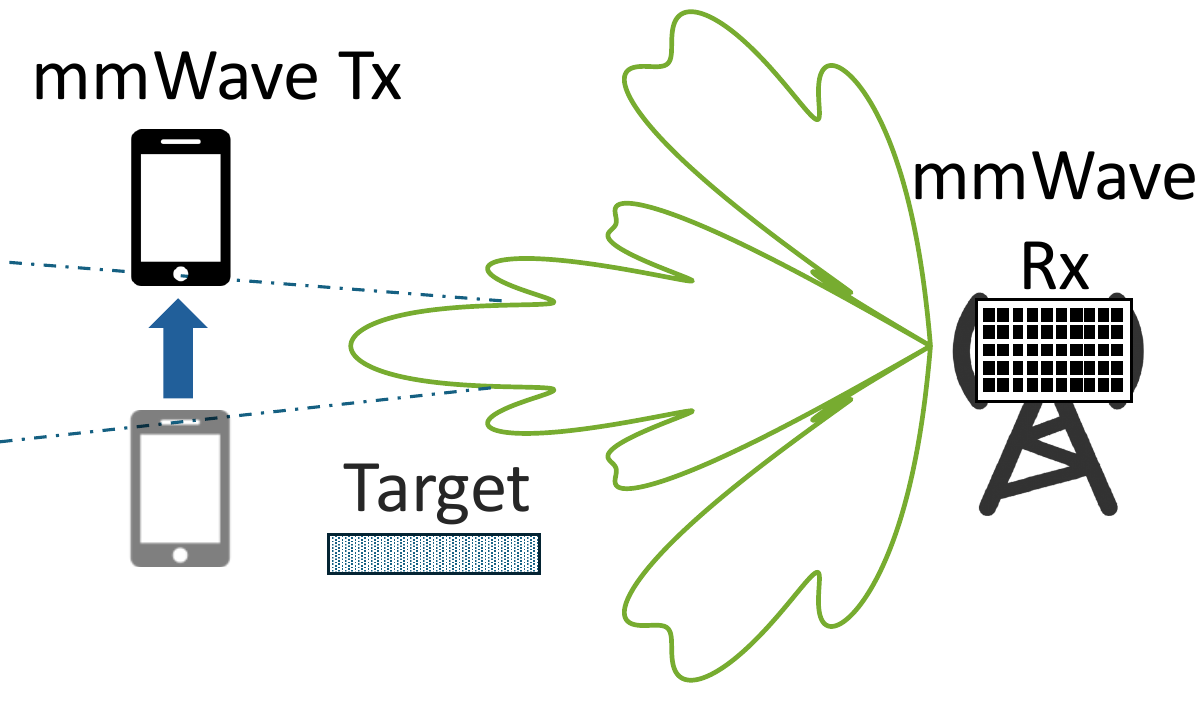}                  \vspace{-6mm}

         \caption{}
         \label{fig:micro_TxMobility_scene}
     \end{subfigure}\hspace*{\fill}
     \begin{subfigure}{0.24\textwidth}
         \centering
         \includegraphics[width=\textwidth]{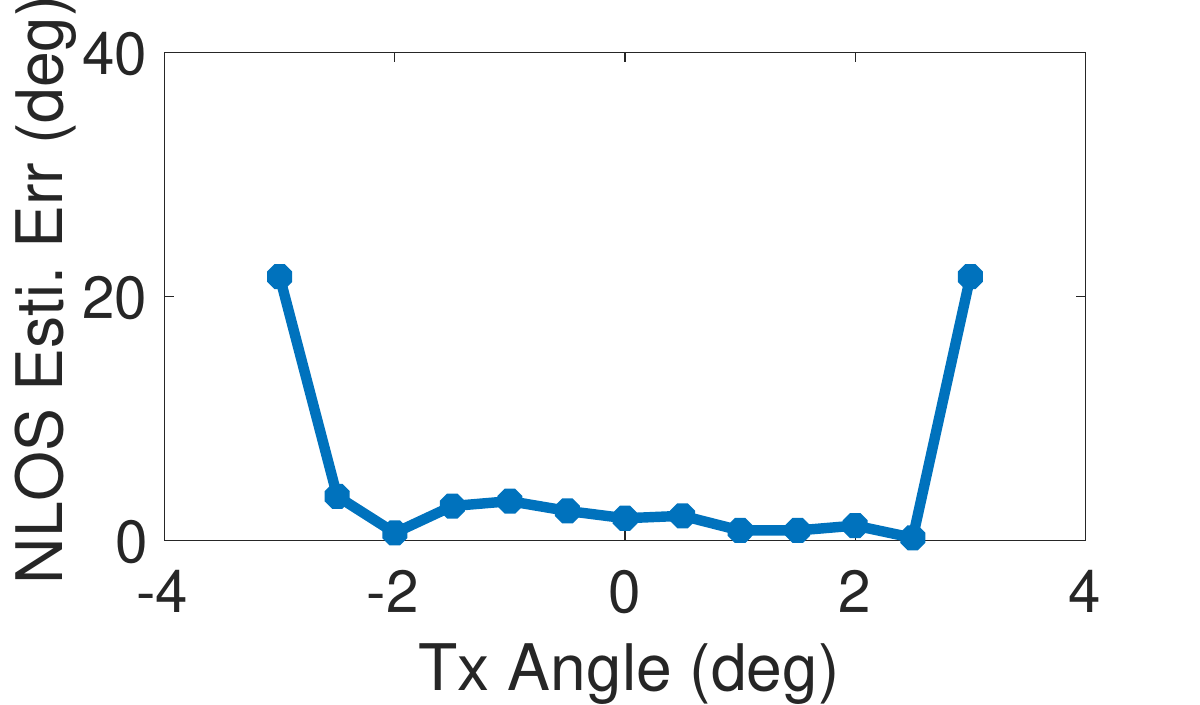}
                  \vspace{-6mm}

         \caption{}
         \label{fig:micro_TxMobility_result}
     \end{subfigure}
             \vspace{-6mm}
    \caption{The impact of slight nodal mobility without beam adaptation: (a) Schematic of the setup; (b) NLOS angle estimation error.}
    \vspace{-4mm}
    \label{fig:micro__TxMobility}
\end{figure}

Obviously, when the TX moves outside of the RX's beamwidth, the SNR drops significantly and the communication link is compromised. In this case, \acrshort{design} that enables sensing on top of data reception cannot be implemented. Indeed, according to IEEE 802.11ay and 3GPP, a beam re-alignment procedure should be initiated in this scenario. 
However, the key question is whether \acrshort{design} is robust to slight nodal mobility that does not trigger beam re-alignment protocol. To test this, we move the TX from -3$\degree$ to 3$\degree$ relative to the RX in 0.5$\degree$ step as depicted in Fig. \ref{fig:micro_TxMobility_scene}. The receiver mainlobe is steered towards the broadside (i.e., 0$\degree$).  

The result is shown in Fig. \ref{fig:micro_TxMobility_result}. We observe that the NLOS angle estimation error remains low in the angular range of -2.5$\degree$ to 2.5$\degree$, and grows rapidly beyond that. This is because as the transmitter moves out of the RX beamwidth, the spatial separation between the two functionalities (communication and sidelobe sensing) does not hold anymore, i.e., the significant interference of the LOS communication link hinders obtaining accurate NLOS fingerprints through sidelobe perturbations. Unsurprisingly, due to TX motion, the communication SNR also drops which would trigger beam re-alignment protocols. Once the new directional beam is established, \acrshort{design} will be able to provide accurate sidelobe sensing on top of the underlying communication link. Therefore, \acrshort{design} is resilient to slight TX motion as long as the transmitter does not fall outside of the receiver half-power beamwidth, in which both communication and sensing performances are compromised.


\subsubsection{Mobile Reflector Trajectory}
So far, we have configured a reflector moving perpendicular to the TX-RX LOS path, similar to Trace 1 in Fig.~\ref{fig:app_floorplan}. Here, we argue that \acrshort{design} is robust to other arbitrary trajectories that a reflector might take as long as the object's reflection footprint can be observed at the receiver. Hence, we place the TX and RX at a 3 m distance in a conference room and transmit 4QAM data symbols at a rate of 25 MHz. We configure three different trajectories as shown in Fig. \ref{fig:app_floorplan}. 
The estimation results are depicted in Fig. \ref{fig:app_trace1}, \ref{fig:app_trace2} and \ref{fig:app_trace3}, respectively. We observe that the ground truth angles match well with angles estimated by \acrshort{design}. The error is higher around 0$\degree$ for all three traces as the reflector is partially blocking the LOS path. Further, the error increases as the reflector approaches the FoV of TX or RX antenna arrays. Overall, we observe that the NLOS sensing accuracy is oblivious to the distance of the reflector from the TX or RX. 

\begin{figure}[t]
     \centering
     \medskip \hspace*{0cm}
     \begin{subfigure}{0.24\textwidth}
         \centering
         \includegraphics[width=\textwidth]{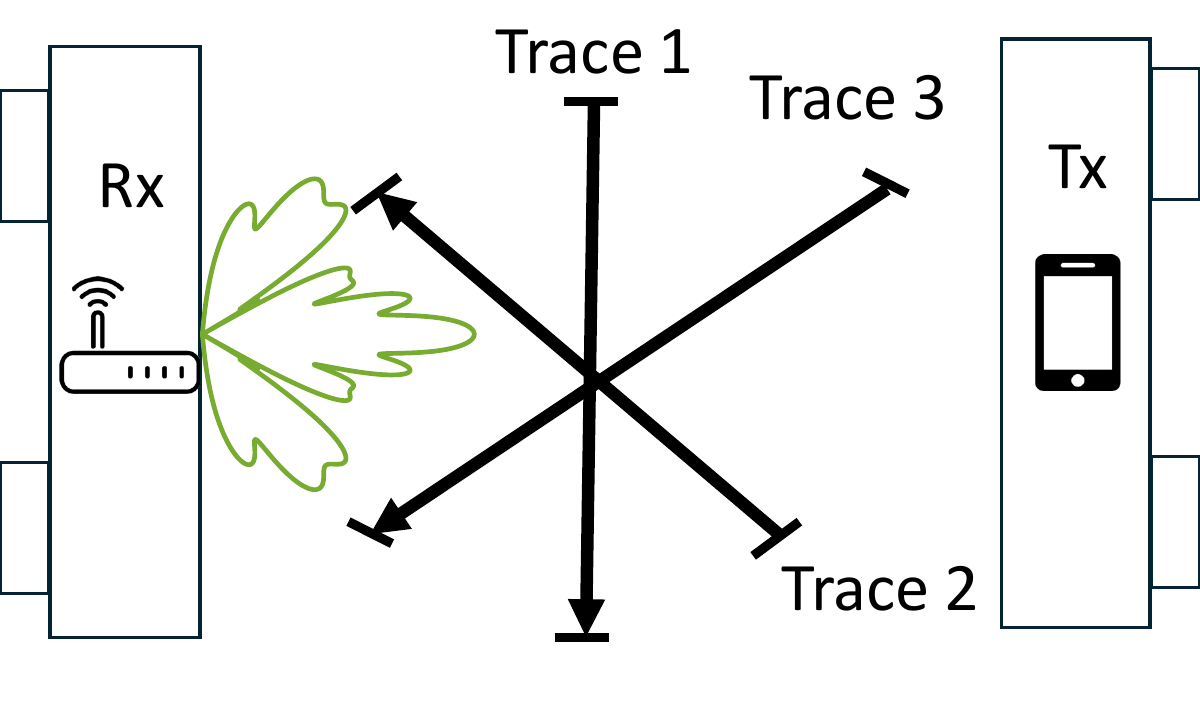}
                  \vspace{-5mm}
         \caption{Experiment Floor Plan.}
         \label{fig:app_floorplan}
     \end{subfigure}\hspace*{\fill}
     \begin{subfigure}{0.24\textwidth}
         \centering
         \includegraphics[width=\textwidth]{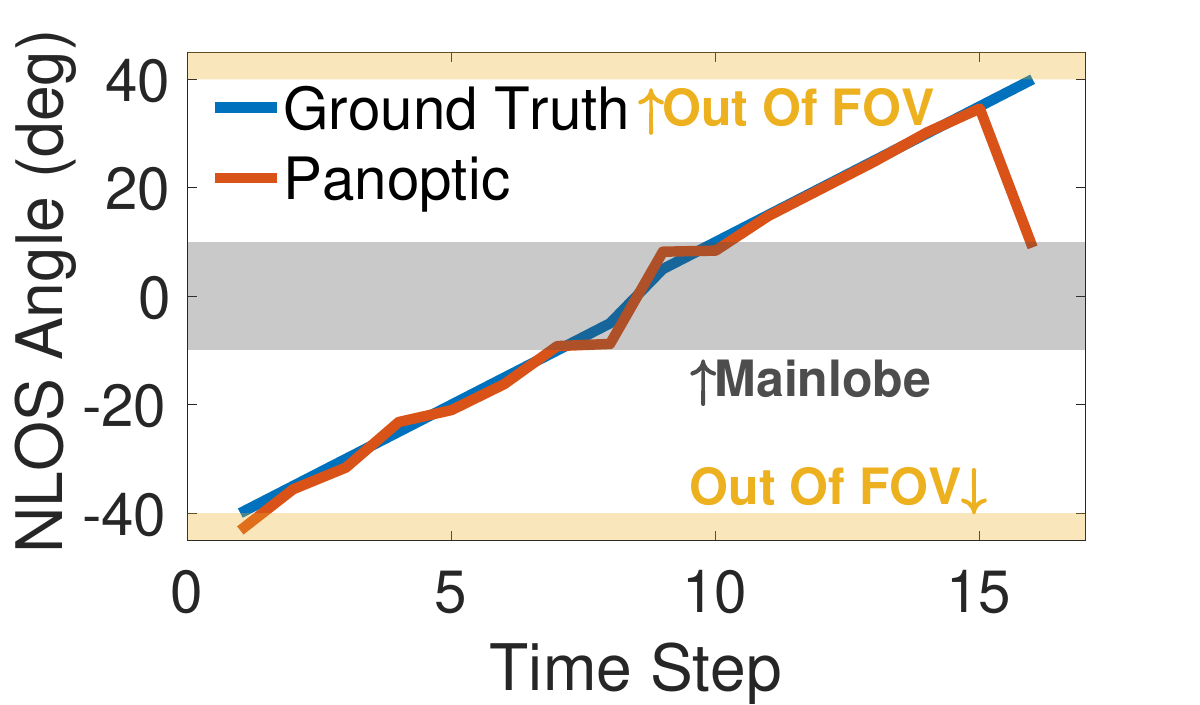}
                           \vspace{-5mm}
         \caption{Trace 1}

         \label{fig:app_trace1}
     \end{subfigure}
     \medskip \hspace*{0cm}
     \begin{subfigure}{0.24\textwidth}
         \centering
         \includegraphics[width=\textwidth]{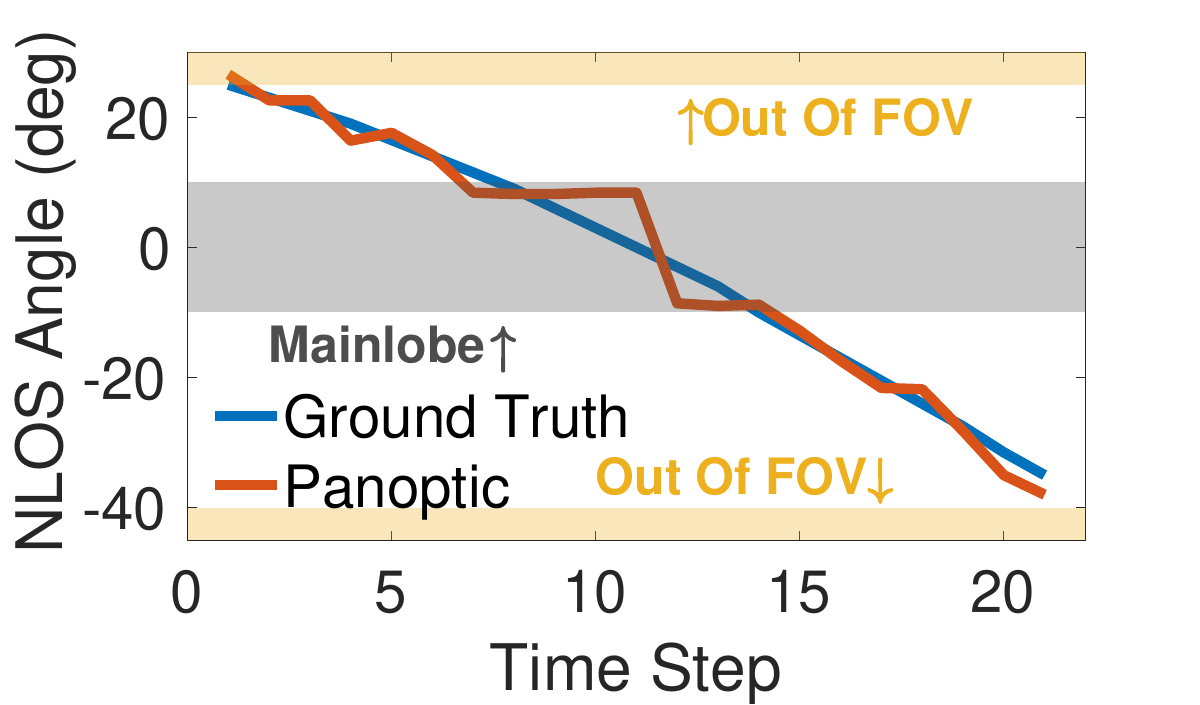}
                  \vspace{-5mm}
         \caption{Trace 2}
         \label{fig:app_trace2}
     \end{subfigure}\hspace*{\fill}
     \begin{subfigure}{0.24\textwidth}
         \centering
         \includegraphics[width=\textwidth]{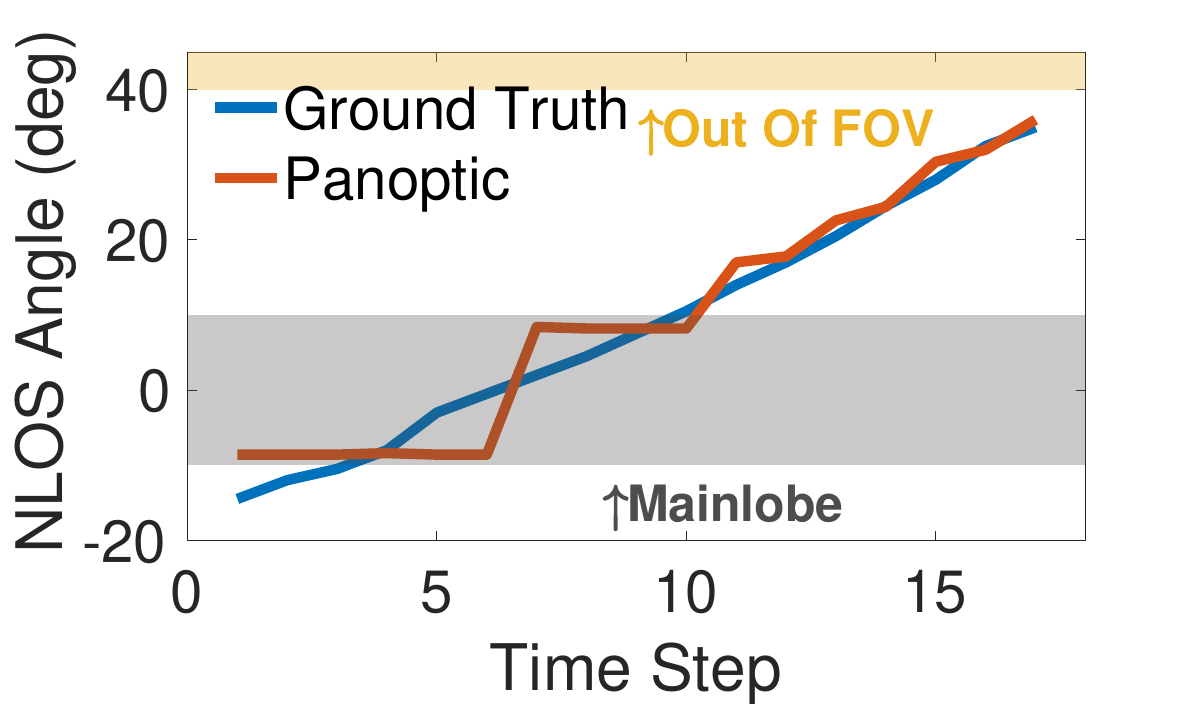}
                  \vspace{-5mm}
         \caption{Trace 3}
         \label{fig:app_trace3}
     \end{subfigure}
             \vspace{-7mm}

    \caption{\acrshort{design} with a different target trajectory.}
        \vspace{-7mm}
        \label{fig:trace}
\end{figure}

Finally, \textcolor{black}{it is worth mentioning that while   we adopt a relatively low antenna switching rate for proof of concept experiments,  it is possible to increase the antenna switching rate in practice for faster environmental tracking.} With a simple calculation, we find that the antenna switching rate needed to angularly track mobile targets is well below the capability of today's off-the-shelf arrays, allowing for real-time estimation and angular tracking in practical mobile environments. Specifically, given a symbol rate of $f_s$ and the number of data symbols per beam $N$, the beam switching rate can be found as $\frac{f_s}{N}$. For instance, with a symbol rate of $1$ GHz and minimum required $N=750$ (for reliable NLOS signature extraction as discussed in Sec.~\ref{subsec:env_exp}), we reach the beam switching rate of $1.33$ MHz, which is an order of magnitude smaller than the Sivers array switching rate of $18.18$ MHz. Indeed, previous work showed an antenna switching rate of about $18$ MHz on Sivers arrays with FPGA \cite{lacruz2020mm}. Note that supporting higher symbol rates is trivial, as one can consider increasing $N$, which helps with better averaging and NLOS signature extraction from weak sidelobe perturbations. Further, as shown by our experimental results in Sec.~\ref{subsec:env_exp}, testing with $M=100$ different random sidelobe configurations is sufficient for accurate NLOS sensing. Hence, the sensing frequency is calculated by $\frac{f_s}{N\times M}=13.3$ KHz, following the previous example. Such a high sensing rate is sufficient to angularly track even high-speed targets \textcolor{black}{in dynamic environments}, e.g., cars. \textcolor{black}{Specifically, at this rate, even a fast-moving object such as a car traveling at 100 km/h (27.8 m/s) moves only about 2.1 mm during the sensing intervals, which is negligible and would not yield any meaningful AOA variations. 
}

\begin{figure}[t!]
     \medskip
     \centering
    \includegraphics[width=0.36\textwidth]{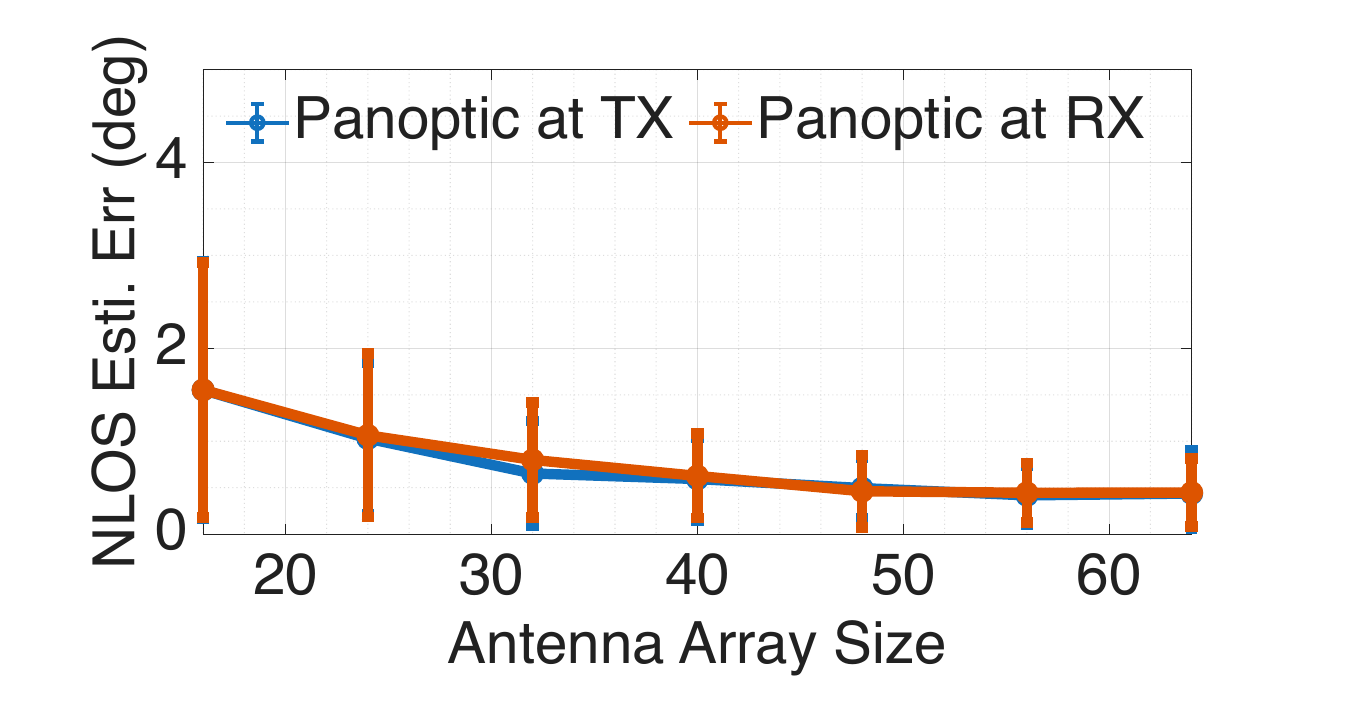}
                  \vspace{-1mm}
     \vspace{-1mm}
    \caption{\textcolor{black}{ \acrshort{design} implementation on TX-Side vs RX-side: Creating compressive sidelobes at the TX yields the same performance as implementing it on the RX side.}}
     \vspace{-3mm}
     \label{fig:micro_tx_rx_nlos_err}
\end{figure}

\begin{figure}[t!]
     \medskip
     \begin{subfigure}{0.24\textwidth}
         \centering
         \includegraphics[width=\textwidth]{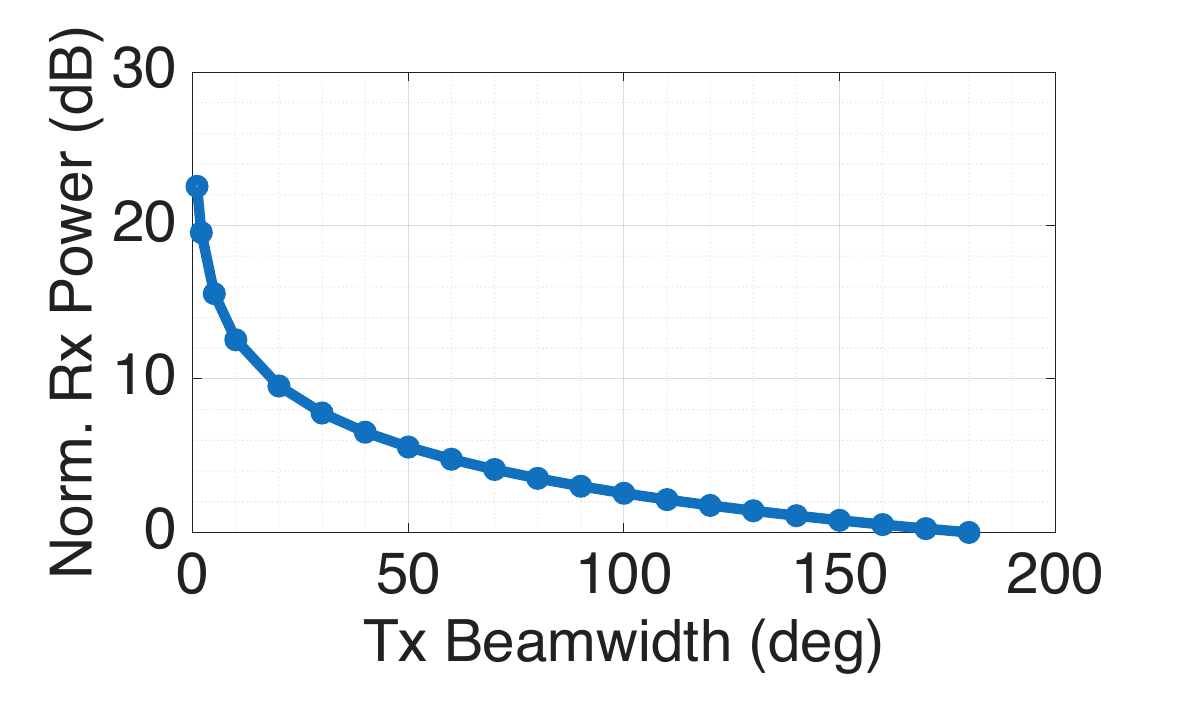}
                  \vspace{-6mm}
         \caption{}
         \label{fig:micro_comm_power_tx_ant_num}
     \end{subfigure}\hspace*{\fill}
     \begin{subfigure}{0.24\textwidth}
         \centering
         \includegraphics[width=\textwidth]{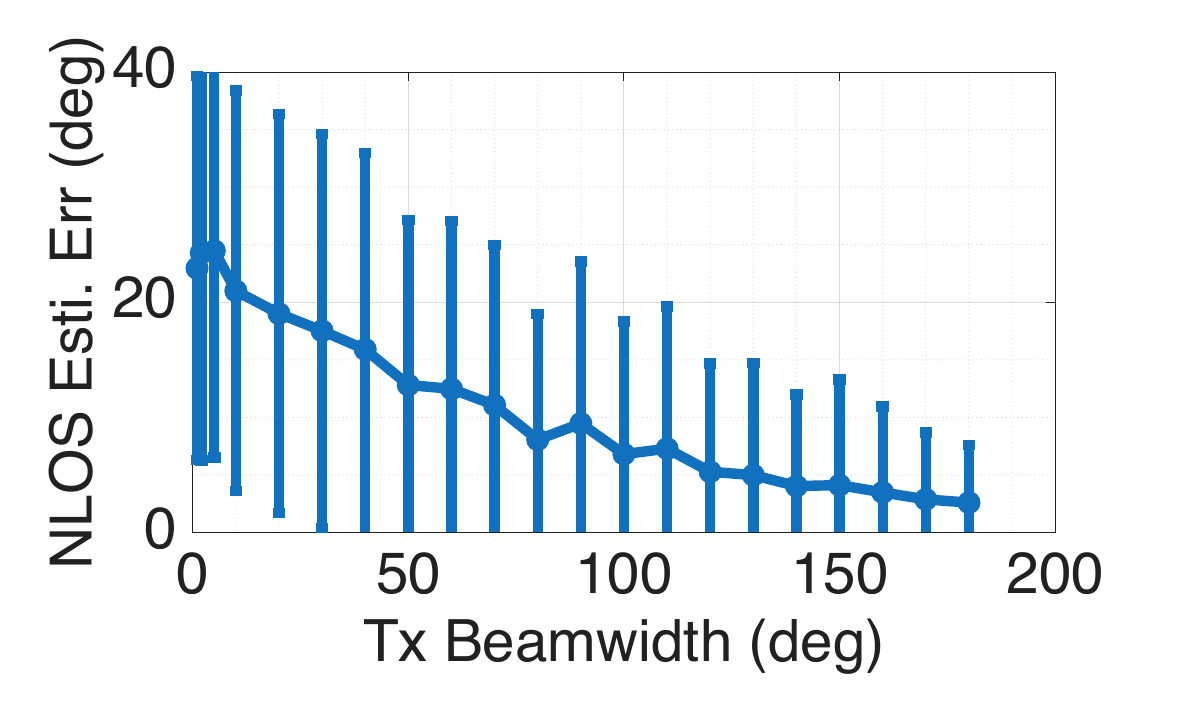}
                  \vspace{-6mm}

         \caption{}
         \label{fig:micro_tx_ant_num_nlos_err}
     \end{subfigure}
     \vspace{-6mm}
    \caption{\textcolor{black}{The tradeoff in the TX beamwidth:  (a) wider TX beams decreases the received power for data communication; (b) wider TX beams increase the amount of power illuminated on the reflector, hence improving sensing performance.}}
     \vspace{-5mm}
\end{figure}

\textcolor{black}{
\subsubsection{\acrshort{design} Implementation on TX-Side vs RX-side}
Here, we attempt to answer a simple question: What if \acrshort{design} is adopted at the transmitting beams instead of receiving beams? To answer this question, we conduct simulations in which we consider different array sizes on one end (TX or RX) while the other end has a simple omnidirectional beam.  We set 189 random compressive sidelobe beams with 4 antennas turned off, regardless of the array size. The distance between TX and RX is set at 2 meters. 50 randomly located reflector configurations are considered, such that they are all within +/- 45 degrees of the TX/RX nodes and the reflectors create a specular NLOS path between the TX and RX.  The NLOS angle estimation result is shown in Fig. \ref{fig:micro_tx_rx_nlos_err}. We observe that \acrshort{design}'s sensing performance is not a function of the TX or adoption, but instead, it depends on the number of random beam configurations that are tested.  The slight improvement with larger arrays comes from a greater number of distinct sidelobe patterns. As array size increases, the space of possible OFF positions expands, reducing sidelobe correlation and improving compressive sensing performance.}

\color{black}
\subsubsection{Impact of the TX Beamwidth on Communication and Sensing Performance}

We also explore how the TX beamwidth affects both communication and sensing performance, assuming a fixed TX beam and the Rx adopting compressive sidelobe beams. In our simulation, we vary the TX beamwidth from 1$\degree$ to 180$\degree$, and use a 16-element Rx antenna array with 4 elements turned OFF. We generate 189 random compressive sidelobe beams, with the TX and RX spaced 2 m apart. We consider 200 random reflector locations, all within $\pm45\degree$ of the nodes. Each reflector creates a specular NLOS path between the TX and RX. The result is shown in Figs. \ref{fig:micro_comm_power_tx_ant_num} and \ref{fig:micro_tx_ant_num_nlos_err}. As the Tx beam becomes wider, the communication power decreases because the energy is spread over a broader area. On the other hand, the NLOS sensing accuracy improves since more of the surrounding space is illuminated, and reflections from the environment are more likely to be captured. In practice, an adaptive algorithm can be used to adjust the TX beamwidth depending on different application scenarios that vary in the required link capacity, SNR range, environment complexity, etc., balancing between communication and sensing performance.

\color{black}



\textcolor{black}{\subsection{Sensing Multiple Reflectors: Coherent vs Non-Coherent}} \label{subsec:cohe_wide_OFDM}
In previous experiments, we evaluated the JCS capabilities of \acrshort{design} when there is one dominant reflecting target in the environment. Here, we show that \acrshort{design} can be used to extract multiple targets. Further, for comparison purposes, we also implement \acrshort{design} under the coherent setting, where complex NLOS signatures (amplitude and phase) for various random sidelobe patterns are used for NLOS sensing. 

\textbf{Setup.}  We use two reflectors of the same size $0.3 \times 0.3$ m for this setup. \textcolor{black}{We place one reflector at $10\degree$ relative to the LOS angle and move another reflector from -35$\degree$ to 35$\degree$ with 5$\degree$ step size, skipping  0$\degree$. We send modulated data symbols (4QAM) with OFDM waveforms under each configuration. We switch the sidelobe pattern every two data symbols at a high rate of 2.5 MHz.  We collect the data-modulated signals under  60 beams, and evaluate \acrshort{design} under both coherent and non-coherent assumption as described in Sec.\ref{subsubsec:ref_sense_framework}. We also collect data from non-coherent 189 beams to compare with previous results. We repeat all experiments 50 times.}

\textbf{Result.} First, we observe that the BER does not get negatively impacted by the placement of two reflectors (not shown), i.e., 0 out of  \textcolor{black}{6.39$\times 10^7$ bits, even in a wideband OFDM setting}. Further, Fig. \ref{fig:micro_MultiRef_GT_Angle} shows the empirical cumulative distribution function (CDF) of error across all configurations. First, as expected, assessing more pseudorandom sidelobe configurations help with improving NLOS estimation under the non-coherent setting. But interestingly integrating phase and amplitude information provides a better AoA estimation performance  compared with  amp-only sensing framework even with more sidelobe configurations.  Fig. \ref{fig:micro_MultiRef_Diff_Angle} confirms that the error is relatively consistent independent of the reflector location.  Indeed, the mean angle error (for both reflectors) is measured at 1.70$\degree$ using the coherent scheme, and the standard deviation remains low at 0.93$\degree$. The $90^{th}$ percentile is 3.15$\degree$. Hence, our results confirm that \acrshort{design} can successfully extract multiple targets in the environment while detecting the unknown transmitted  OFDM data symbols.

\vspace{+3mm}

\begin{figure}[t!]
     \medskip
        \begin{subfigure}{0.48\textwidth}
         \centering
         \includegraphics[width=\textwidth]{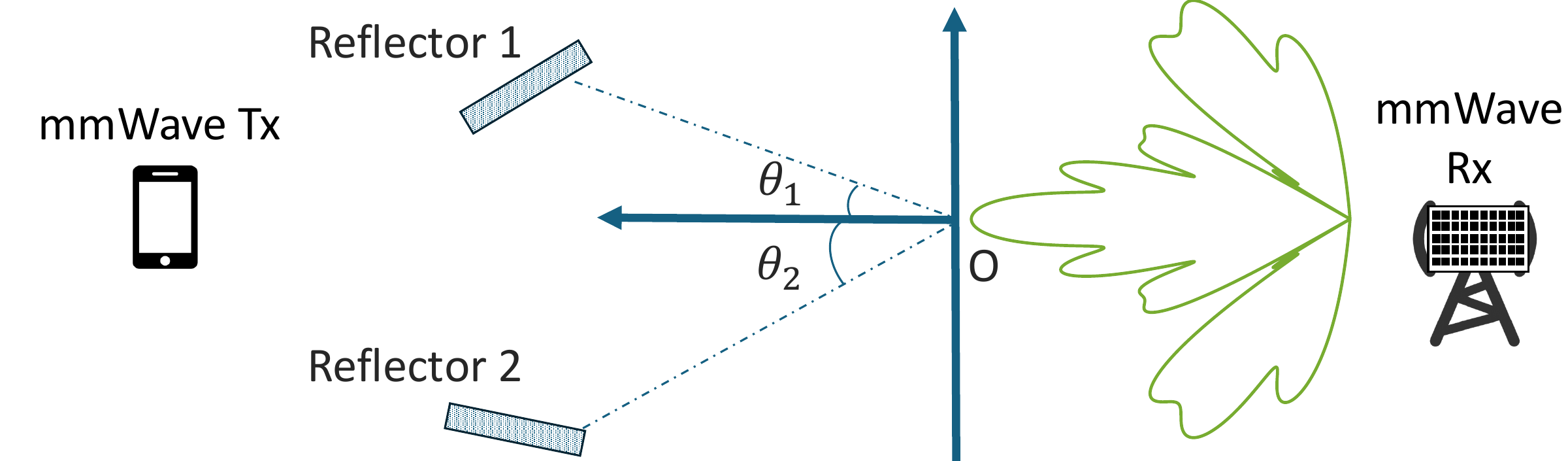}
         \vspace{-7mm}
         \caption{}
         \label{fig:micro_MultiRef_scene}
     \end{subfigure}
     \begin{subfigure}{0.24\textwidth}
         \centering
         \includegraphics[width=\textwidth]{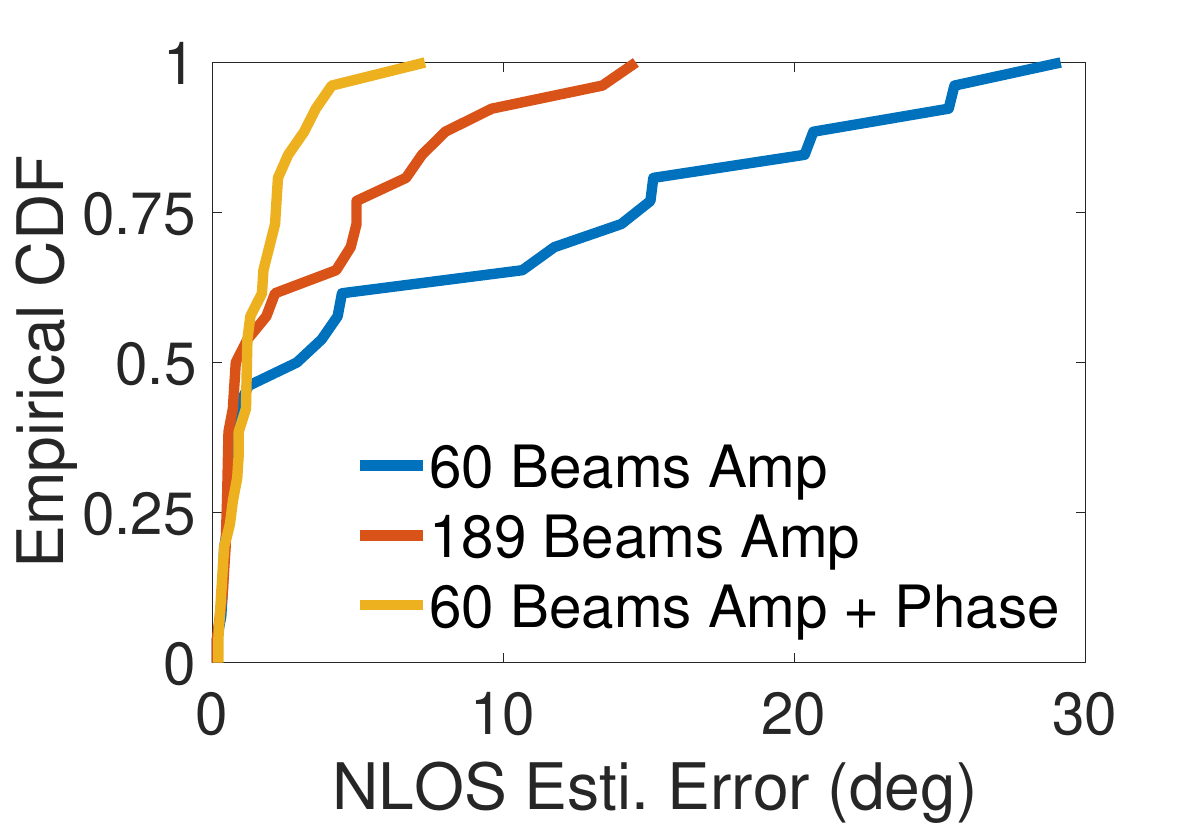}
         \vspace{-6mm}
         \caption{}
         \label{fig:micro_MultiRef_GT_Angle}
     \end{subfigure}\hspace*{\fill}
     \begin{subfigure}{0.24\textwidth}
         \centering
         \includegraphics[width=\textwidth]{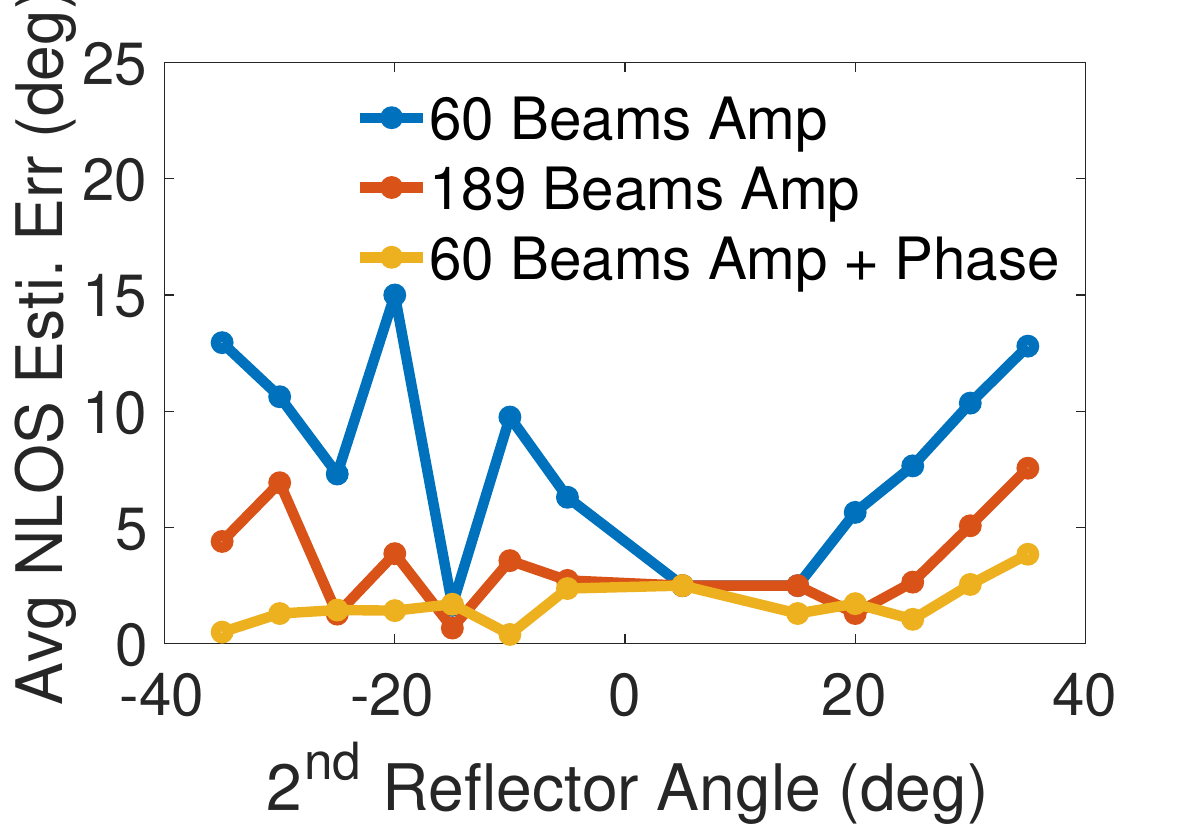}
         \vspace{-6mm}
         \caption{}
         \label{fig:micro_MultiRef_Diff_Angle}
     \end{subfigure}
     \vspace{-6mm}
    \caption{\acrshort{design} supports sensing multiple reflectors: (a) Schematic of the setup; (b) Empirical CDF of error; (c) AoA error as a function of the angular separation between NLOS paths.}
    \label{fig:micro_MultiRef}
    \vspace{-4mm}
\end{figure}

\section{Discussion} \label{sec:discussion}
\textbf{Super-Resolution Sensing with \acrshort{design}.} 
While this paper builds a novel foundation for joint communication and sensing using antenna subset switching, there are several schemes that can be adopted in future work to improve the sensing resolution. \textcolor{black}{First, combining phase and amplitude readings, one can extend \acrshort{design} to more advanced sensing tasks like environmental imaging, human activity recognition, etc. We leave this exploration for future work.}  \textcolor{black}{Doppler extraction, which offers information about the velocity of objects in the environment, is left for future work. Extracting Doppler shifts requires coherent phase measurements across successive transmissions, which in turn demands tighter synchronization between nodes and stable local oscillators—challenges that are more pronounced in mmWave systems. With Doppler information, our system could capture dynamic environmental changes, enabling several new applications.} 

Second, \acrshort{design} chooses the antenna subset randomly without any feedback from the RX. In principle, not every sidelobe perturbation pattern provides the same amount of information for a particular target extraction. In the future, we will explore adaptive compressive sidelobe selection strategies. Last but not least, more advanced data processing methods, such as data-driven and ML-assisted techniques, could potentially improve sensing accuracy. \textcolor{black}{ Machine learning-assisted methods hold strong potential for improving sensing accuracy due to their ability to capture nonlinear features and robustness to noise. In the current approach, Panoptic builds a simulated library of NLOS signatures and uses correlation to estimate angles by matching received signatures to this library. This process could be enhanced using sequence-based machine learning models. Specifically, the sequence of measurements from different compressive sidelobe beams can be treated as input, with the goal of predicting the angle of arrival. Well-established models like RNNs or Transformers are well-suited for this task. A practical strategy would involve pretraining these models on simulated data to capture the structure of \acrshort{design}’s algorithm, followed by fine-tuning on a small set of real-world measurements to adapt to hardware imperfections and environmental nuances. We leave this for future work. }

\textbf{Scaling to Large Arrays .} The mmWave nodes we use in evaluating \acrshort{design} have 16 antenna elements on the azimuth plane. Indeed, mmWave AP/BS (and even mmWave handheld devices) are often equipped with larger antenna arrays. Larger antenna arrays expand the randomness space for sidelobe perturbations and result in better angular localization resolutions. In this case, turning off a small number of antennas can provide sufficient random space for sensing; hence, the penalty of compressive sidelobe transmission on the communication SNR and BER becomes negligible. As discussed in Sec.~\ref{subsubsec:ref_sense_framework}, the number of compressive sidelobe beam measurements only grows with $O({\rm log} N)$, where $N$ is the number of antennas in array; hence, \acrshort{design} is scalable to large arrays. 

\textbf{Integrating Inputs from Preambles and Data Symbols.} While past work exploited known signals (e.g., preambles) for channel sensing, this paper introduced a paradigm shift in JCS systems by jointing detecting channel components and unknown data symbols. A frame includes short preambles and many data symbols. Hence, in principle, one can leverage inferences from preambles as known reference signals (albeit available at lower duty-cycle) to further improve sensing resolution in \acrshort{design}. We will leave this for future exploration. 

\textbf{Extension to 3D Imaging.} In this paper, we have implemented 2D sensing using mmWave arrays that support azimuth beamforming only. In principle, with radios supporting beamforming in both azimuth and elevation, \acrshort{design} can achieve 3D imaging. While there are numerous applications for such joint communication and 3D imaging framework, mapping the sidelobe perturbations in azimuth and elevation to 3D fingerprints requires further exploration that we leave for future work.

\textcolor{black}{\textbf{LOS Blockage and  Compatibility with Existing Protocols.} When the direct path is blocked or unavailable for communication, \acrshort{design} relies on the strongest NLOS path for communication as suggested by conventional protocols . In such a case, we can detect other NLOS paths in the same way as discussed in the paper . \acrshort{design} is compatible with existing protocols like IEEE 802.11ad and 3GPP by replacing directional beams in data transmission with compressive sidelobe beams, which enables reflector sensing during communication at the same time, frequency and using the same hardware resources.}

\section{Conclusion} \label{sec:conclusion}
This paper presents \acrshort{design}, a novel system architecture for integrated mmWave communication and sensing that share time, frequency, and hardware resources. \acrshort{design}'s foundation lies on a unique beam manipulation technique realized by antenna subset modulation using a single RF-chain mmWave array. We introduce compressive sidelobe radiation, a novel beam pattern that simultaneously supports a stable high-gain communication link as well as pseudorandom sidelobe perturbations enabling NLOS sensing. Our extensive over-the-air experiments using commercial 60 GHz arrays demonstrate a mean NLOS angle estimation error of below 1.3$\degree$ without compromising accurate symbol extraction. This work paves the way for true joint communication and sensing in mmWave wireless networks. 

\section*{Acknowledgment}
This work was supported in part by the National Science Foundation (grant 2148271) and by
The Center for Ubiquitous Connectivity (CUbiC), sponsored by Semiconductor
Research Corporation (SRC) and Defense Advanced Research
Projects Agency (DARPA) under the JUMP 2.0 program.

\bibliographystyle{IEEEtran}
\bibliography{bib/IEEEabrv,bib/IEEEexample}

\end{document}